\documentclass[preprint,times,12pt]{elsarticle}
\usepackage[top=3.5cm,bottom=3.5cm,left=3.0cm,right=3.0cm]{geometry}
\usepackage{lineno,hyperref}
\modulolinenumbers[0]

\usepackage{amsmath}
\usepackage{amsfonts}
\usepackage{amssymb}
\usepackage{amsxtra} 
\usepackage{mathtools}
\usepackage[OMLmathsfit]{isomath}
\usepackage{bm}                          
\usepackage{float}
\usepackage[version=4]{mhchem}
\usepackage{caption}
\usepackage{subcaption}
\usepackage{siunitx}
\usepackage{hyperref}
\usepackage[capitalise]{cleveref}
\usepackage{graphicx}
\usepackage{libertine}
\usepackage[explicit]{titlesec}
\usepackage{csquotes}
\usepackage[table,xcdraw]{xcolor}
\definecolor{darkgreen}{rgb}{0.0,0.4,0.0}

\begin{document}

\begin{frontmatter}
\title{Chemo-Mechanical Phase-Field Modeling of Iron Oxide Reduction with Hydrogen}

\author[MPIE]{Yang Bai\corref{mycorrespondingauthor}}
\ead{y.bai@mpie.de}

\author[MPIE]{Jaber Rezaei Mianroodi\corref{mycorrespondingauthor}}
\ead{j.mianroodi@mpie.de}

\author[MPIE]{Yan Ma}
\author[MPIE]{Alisson Kwiatkowski da Silva}
\author[MPIE,RWTH]{Bob Svendsen}
\author[MPIE]{Dierk Raabe}

\address[MPIE]{Microstructure Physics and Alloy Design, Max-Planck-Institut f\"ur Eisenforschung, D\"usseldorf 40237, Germany}

\address[RWTH]{Material Mechanics, RWTH Aachen University, Schinkelstr. 2, 52062 Aachen, Germany}

\cortext[mycorrespondingauthor]{Corresponding author}

\begin{abstract}
The reduction of iron ore with carbon-carriers is one of the largest sources of greenhouse gas emissions in the industry, motivating global activities to replace the coke-based blast furnace reduction by hydrogen-based direct reduction (HyDR).
Iron oxide reduction with hydrogen has been widely investigated both experimentally and theoretically.
The process includes multiple types of chemical reactions, solid state and defect-mediated diffusion (by oxygen and hydrogen species), several phase transformations, as well as massive volume shrinkage and mechanical stress buildup.
However, studies focusing on the chemo-mechanical interplay during the reduction reaction influenced by microstructure are sparse.
In this work, a chemo-mechanically coupled phase-field (PF) model has been developed to explore the interplay between phase transformation, chemical reaction, species diffusion, large elasto-plastic deformation and microstructure evolution.
Energetic constitutive relations of the model are based on the system free energy which is calibrated with the help of a thermodynamic database.
The model has been first applied to the classical core-shell (w\"ustite-iron) structure. Simulations show that the phase transformation from w\"ustite to $\alpha$-iron can result in high stress and rapidly decelerating reaction kinetics.
Mechanical stresses can contribute elastic energy to the system, making phase transformation difficult. Thus slow reaction kinetics and low metallization are observed.
However, if the stress becomes comparatively high, it can shift the shape of the free energy from a double-well to a single-well case, speed up the transformation and result in a higher reduction degree than the lower stress case.
The model has been further applied to simulate an actual iron oxide specimen with its complex microstructure, characterized by electron microscopy.
The experimentally observed microstructure evolution during reduction is well predicted by the model.
The simulation results also show that isolated pores in the microstructure are filled with water vapor during reduction, an effect which influences the local reaction atmosphere and dynamics.
\end{abstract}


\end{frontmatter}

\section{Introduction}
Global warming and efforts to reduce its impact are one of the greatest current challenges.
The production and use of iron and steel is one of the most important cornerstones of our civilization and technology, but it is also one of the largest contributors to the greenhouse gas emissions, due to the redox reaction~\ce{Fe2O3 + 3CO -> 2Fe + 3CO2}, which is behind current reduction technologies.
Producing one ton of steel generates between 1.85 and 2.1 tons of carbon dioxide~\cite{pineau2006kinetics,patisson2020hydrogen}. In terms of annual global production, the 1800 million tons (Mt) of steel produced in 2019 dwarfs that of aluminum, with only 94 Mt the second most common metal produced.
Global steel production on this scale is responsible for about 6\% of the global energy consumption and for almost 25\% of the industrial~\ce{CO2}-equivalent emissions~\cite{Raabe2019}.
Therefore, reducing the emission of greenhouse gases has become one of the most essential topics in the manufacturing sector.
In the search for alternatives to carbon monoxide, hydrogen, as one of the buffer molecules for storing and using sustainable energy, becomes attractive in that context.
Compared with the current carbon-based iron ore reduction, green hydrogen is a more environmentally friendly reducing agent since the reaction product of the underlying redox reaction is water.
It can also replace the expensive metallurgical coke production step and eliminate the subsequent decarburization steps in steel production, where the near eutectic Fe-C 'pig-iron' alloy, produced by blast furnaces, is converted into low-C steels~\cite{pineau2006kinetics}.
Therefore, Hydrogen-based Direct Reduction of Iron oxides (HyDRI), instead of the use of carbon monoxide as the reducing gas, is a promising way to drastically reduce greenhouse gas emissions in steel production, thus attacking this grand challenge with advanced technologies, based on a clear understanding of the underlying physical and chemical fundamentals.

Iron oxide (in form of hematite) is typically reduced in three stages in HyDRI: hematite (\ce{Fe2O3}) to magnetite (\ce{Fe3O4}), magnetite (\ce{Fe3O4}) to w\"ustite (\ce{FeO}), and w\"ustite (\ce{FeO}) to sponge iron, i.e. highly porous Fe.
This sequence is thus characterized by several phase transitions, each associated with the oxygen mass loss and volume change, until the last step to pure iron, from w\"ustite to $\alpha$-iron.
Thus, the HyDRI process is characterized by a complex chemo-mechanical interplay of the different mechanisms involved, specifically the reactions, mass transport, and volume changes.
Among the reaction steps, w\"ustite reduction to iron is the slowest one by nearly an order of magnitude lower reaction kinetics compared with the other two steps~\cite{KIM2021116933}, therefore, it plays an important role in determining the overall rate of the reactions~\cite{feinman1999direct,ghosh2008iron,KIM2021116933}.
Although most aspects of carbon- and hydrogen-based DRI are similar~\cite{john1984breakdown,john1984establishment,john1982microstructural,nicolle1979mechanism}, some significant differences must be underlined.
For example, the hydrogen reduction is generally endothermic, whereas the carbon monoxide reduction is exothermic.
Above 800~\si{\degreeCelsius}, however, thermodynamics are more favorable with hydrogen than with carbon monoxide, where the reduction rate with~\ce{H2} is much higher than the case with~\ce{CO} at 850~\si{\degreeCelsius}~\cite{zuo2015reduction,spreitzer2019reduction}.
Furthermore, morphological analysis of reduced iron oxide samples (\ce{Fe2O3}) by~\ce{H2} reveals aggregation of the reaction product (compact iron layer), which is not seen in oxides reduced by~\ce{CO} at temperatures above  420~\si{\degreeCelsius}~\cite{pineau2006kinetics}.
In the study based on thermo-gravimetry,~\citet{Kuila2016} demonstrated that the utilization of hydrogen (10.68~\si{\mole}) is more effective than carbon monoxide (12~\si{\mole}) in the reduction of 1~\si{\mole} of magnetite (\ce{Fe3O4}) ore fines. 
As a result, the interaction between the reducing gases and w\"ustite affects the overall efficiency of a direct reduction reactor in terms of both thermodynamics and microstructure morphology.
Therefore, w\"ustite reduction is extremely important in the majority of commercially used gas-based direct reduction processes for producing sponge iron.
In HyDRI processes,~\ce{FeO} plays an important role as an intermediate reduction product~\cite{feinman1999direct}.
A more detailed understanding of the interplay of the many interacting physical and chemical phenomena, such as hydrogen and oxygen diffusion, phase transformation, mechanical stress buildup and the associated (in)elastic deformation effects, the positions where water is formed and trapped, as well as crack formation and propagation, during these reduction processes becomes essential.
Motivated by this substantial complexity, several studies have been published to identify and understand the bottleneck effects in HyDRI with respect to the efficiency of the hydrogen gas usage, overall reduction kinetics, and metallization yield~\cite{spreitzer2019reduction,KIM2021116933,baolin2012study,jozwiak2007reduction,mckewan1964influence,moukassi1983study,pineau2006kinetics,sastri1982studies}.

A great deal of previous work has been carried out in order to better understand the interaction and interplay of the physical and chemical processes underlying HyDRI (e.g., kinetics, diffusion, phase transformation, mechanical deformation, pore formation, chemical reaction), as well as their impact on the effectiveness of HyDRI.
With respect to reduction kinetics for example,~\citet{kawasaki1962kinetics} investigated the reaction of iron oxide with carbon monoxide and hydrogen. Their experiments show that counter-diffusion (i.e., diffusion in opposite directions) of reacting and product gases has a significant influence on the reduction rate.
~\citet{sastri1982studies} concluded that differences in the reduction kinetics from pure $\alpha$-\ce{Fe2O3} to doped~\ce{Fe2O3} (mixed with~\ce{Li2O},~\ce{MgO} and other foreign metal oxides) can be attributed to structural factors, instead of electronic factors.
More recently,~\citet{pineau2006kinetics,pineau2007kinetics} has also carried out hydrogen reduction of hematite at low temperatures in the range of 220-680~\si{\degreeCelsius}.
\citet{jozwiak2007reduction} investigated the kinetics of reduction of different iron oxides in hydrogen and carbon monoxide atmospheres at different temperatures.~\citet{barde2016solid} conducted an experimental and numerical study of the thermo-chemical reaction kinetics of HyDRI. 

Besides kinetics, the microstructure of iron-ore and its evolution during reduction have been the focus of a number of investigations~\cite[e.g.,][]{turkdogan1971gaseous,swann1977high,moukassi1983study,rau1987investigation,el1988nucleation,matthew1990situ}.
In particular, the evolution of the phase microstructure during HyDRI is determined by three transformations: (i) hematite to magnetite, (ii) magnetite to w\"ustite (FeO), and (iii) w\"ustite to ferrite (\(\alpha\)-Fe)~\cite[e.g.,][]{zielinski2010reduction}. In turn, these are influenced by the phase microstructure and reduction temperature.
In the early work,~\citet{turkdogan1971gaseous} concluded that the porous nature of iron ore (hematite), as well as the change in pore structure with reduction temperature, have a strong influence on the reduction process.
\citet{swann1977high} observed different types of pore structures in varying proportions depending on the reduction temperature. The temperature and pore network structure influence particularly the transport of reductants to the phase interfaces.
In addition, the transformation of w\"ustite into ferrite at the w\"ustite surface results in a layer of iron between this surface and the reducing gas.
Since the diffusion of reductant through ferrite to the ferrite-w\"ustite interface is slow, this results in a decrease in the rate of the reduction of w\"ustite to ferrite~\cite[e.g.,][]{moukassi1983study}.
The w\"ustite to ferrite transformation results in large volume decrease, for example, a total volume contraction of about 24\% has been reported for iron oxide compacts (\ce{Fe2O3}) reduced by hydrogen~\cite[][]{aa1996effect}. Such a large volume change causes a substantial increase in internal stresses, which in turn influence the phase transformation as well as pores and crack development. 

In addition to these primarily experimental investigations, several theoretical efforts have been made to model aspects of iron-oxide reduction.
Based on the assumption of the shrinking core model, which describes a gradually reducing inner w\"ustite volume inside of a dense iron shell around it, and a quasi steady-state approximation,~\citet{tsay1976modeling} developed a three interface core-shell model (TICSM) for the reduction of hematite disks with a mixture of hydrogen and carbon monoxide.
Later, this model has been successfully applied to predict the reduction degree of the direct reduction process in a shaft furnace arrangement~\cite{tsay1976modeling2}.
\citet{yu1981mathematical} provided a finite element analysis of porous iron oxide pellets under non-topochemical reduction conditions, where multiple chemical reactions may occur in any region of the oxide specimen. Employing this model, the gas profiles for each step can be explicitly calculated.
~\citet{ramachandran1982modeling} critically reviewed recent advances in the modeling of gas-solid non-catalytic reactions, with an emphasis on the TICSM for hematite reduction.
~\citet{negri1991direct} extended the TICSM to investigate the impact of water gas shift reactions on the direct reduction of hematite with hydrogen-carbon monoxide gas mixtures.
More recently, a non-isothermal and non-isobaric mathematical model~\cite{sun1999building} has been developed for the kinetics of iron-ore reduction in an ore/coal composite.
An isothermal TICSM has also been introduced by~\citet{valipour2006mathematical} to simulate the time-dependent kinetic and thermal behavior of a porous iron oxide pellet undergoing chemical reactions with hydrogen, carbon monoxide, and water vapor.
\citet{valipour2007modeling} investigated multiple non-catalytic gas-solid reactions in a moving bed of porous pellets with the help of finite-volume-based numerical modeling.
The same authors~\cite[][]{valipour2009mathematical,valipour2009numerical,valipour2011effect}, also studied the non-catalytic gas-solid reduction reaction with syngas for a hematite pellet and porous w\"ustite. 
The kinetics of HyDRI in a differential micro-packed bed have been modeled recently by~\citet{baolin2012study}.
More recent modeling efforts such as~\cite{xu2013numerical} have begun to extend these earlier reduction models, in particular to three dimensions as well as the inclusion of mechanical effects and their coupling to chemical reactions and kinetics.

Despite their flexibility in predicting the overall reduction degree, certain topological features, and reaction kinetics, the above-mentioned models do not account for several important physical effects.
For instance, the phase transformations from w\"ustite to $\alpha$-iron during the reduction, the entire microstructure evolution of the iron oxide sample with its complex geometry evolution (as observed in experiments~\cite{KIM2021116933,hayes1981microstructural,john1982microstructural}), and the large volume shrinkage of the entire sample as well as phase-specific volume changes and the associated large stresses during the reduction reaction are usually not considered in the modeling of iron-ore reduction.
Motivated by recent experimental observations of the evolution of such complex microstructures and micromechanics during reduction and their effects on reduction kinetics and metallization, we introduce here a chemo-mechanically coupled phase-field (PF) model.
The purpose of the current work is the further extension of these earlier modeling efforts via the inclusion of additional physical mechanisms and processes playing a role in HyDRI process.
These include (i) the phase transformation from w\"ustite to ferrite and evolution of the corresponding phase microstructure, (ii) the large deformation (i.e., volume reduction) due to this transformation, and (iii) internal stresses driving phase transformation and inelastic processes (e.g., dislocation glide).
To this end, finite-deformation phase-field chemo-mechanics for multicomponent, multiphase mixtures~\cite[][]{svendsen2018finite} is employed to formulate the model.
Existing applications of this framework include for example the modeling of lithium ion battery electrodes~\cite[][]{anand2012cahn,di2014cahn,bai2019two,bai2020chemo,santos2020bending}, and nanoscopic defect-solute interaction in engineering alloys~\cite[][]{mianroodi2019atomistic,mianroodi2021phase}. 
Of central importance here is the form of the free energy, whose (local) minimization drives the processes underlying HyDRI, i.e., phase transformations, chemical reactions, and microstructure evolution. 
Chemo-mechanical coupling and the effects of finite deformation due to the transformation of w\"ustite into ferrite are accounted for in this case through the elastic part of this energy.
In addition, the quantitative dependence of the chemical part of the energy in the w\"ustite and ferrite phases on oxygen is determined with the help of the Thermo-Calc TCOX10 database~\cite[][]{andersson2002thermo,sundman1991assessment,hidayat2015thermodynamic}.
The corresponding calibrated model is then applied to the modeling of phase transformation, finite deformation elastoplasticity, and microstructure evolution during HyDRI. 

The work begins with the formulation of the model for HyDRI in~\cref{sec:model} in the framework of finite-deformation phase-field chemo-mechanics.
After discussing model identification/calibration in~\cref{sec:calib}, representative simulation results for HyDRI are presented in \cref{sec:results}.
The work ends with a summary and conclusion in~\cref{sec:conclude}. 
In this work, three-dimensional Euclidean vectors are represented by lower-case \(\bm{a},\bm{b},\ldots\), and second-order Euclidean tensors by upper-case 
\(
\bm{A},\bm{B},\ldots
\), 
bold italic characters. The scalar product of two tensors \(\mathcal{A}\) and \(\mathcal{B}\) of any order is symbolized by \(\mathcal{A}\cdot\mathcal{B}:=A_{i\!j\ldots}B_{i\!j\ldots}\) (summation convention). Further definitions and concepts will be introduced as needed in the sequel. 

\section{Model formulation}
\label{sec:model}
As discussed above, the model formulation is based on finite-deformation phase-field chemo-mechanics in the context of chemical and continuum thermodynamics~\cite[e.g.,][]{Pri54,deG62,Silhavy1997} as well as the mixture theory~\cite[e.g.,][]{Tru84}. 
For simplicity, isothermal and quasi-static mechanical conditions are assumed. Given the solid phases and large deformation, the formulation is referential or "Lagrangian" with respect to the mixture. 
In this case, the densities of all extensive quantities are per unit mixture reference volume. 
During the iron oxide reduction with hydrogen, above 570~\si{\degreeCelsius}, a phase transformation from w\"ustite to $\alpha$-iron occurs.
Since the transformation of w\"ustite to ferrite is nearly one order of magnitude slower than the other transformations discussed above, it represents the rate-limiting process in HyDRI~\cite[][]{feinman1999direct,chatterjee2010sponge,KIM2021116933}.
For simplicity, then, attention is restricted to this transformation here, and the HyDRI process is modeled at 700~\si{\degreeCelsius} in this work.
In HyDRI, H reacts in dissociated form with oxygen (O) at the surface of solid w\"ustite, yielding iron (Fe) and water (\ce{H2O}) as products, i.e., 
\begin{equation}
\ce{FeO}+2\ce{H}\leftrightharpoons\ce{Fe}+\ce{H2O}
\,.
\label{equ:ReaDRI}
\end{equation}
Since Fe is essentially passive here, this reaction is simplified to 
\begin{equation}
\ce{O}+2\ce{H}\ \leftrightharpoons\ \ce{H2O}
\label{equ:ReaIde}
\end{equation} 
in this work. On this basis, the following model is formulated for a mixture of three phases (ferrite, gas, w\"ustite), with diffusing ~\ce{H}, ~\ce{H2O} and ~\ce{O} in all phases.

\subsection{Basic relations}
Assuming the mixture is closed with respect to mass/molar number, the balance relations\footnote{The reduced form~\cref{equ:BalMasComDenMix}${}_{1}$ of component mass balance follows from the general form assuming that the mixture molar number density is constant.} 
\begin{equation}
\dot{x}_{i}
=-\mathop{\mathrm{div}}\bm{j}_{i}
+\sigma_{\!i}
\,,\quad
\bm{0}
=\mathop{\mathrm{div}}\bm{P}
\,,\quad
(\nabla\bm{\chi})
\bm{P}^{\mathrm{T}}
=\bm{P}
(\nabla\bm{\chi})^{\mathrm{T}}
\,,\quad
\skew2\dot{\varepsilon}
=\mathop{\mathrm{div}}
\bm{P}^{\mathrm{T}}
\skew2\dot{\bm{\chi}}
\,.
\label{equ:BalMasComDenMix}
\end{equation} 
for component mass (\(i\in\lbrace\hbox{H, \ce{H2O}, O}\rbrace\)), mixture linear momentum, mixture angular momentum, and mixture energy, respectively, hold in the current case.
In these relations, \(x_{i}\) is the molar number fraction of component \(i\), \(\bm{j}_{i}\) and \(\sigma_{\!i}\) are the corresponding flux and supply-rate densities, respectively. \(\bm{P}\) denotes the mixture first Piola-Kirchhoff (PK) stress, \(\bm{\chi}\) represents the mixture deformation field, and \(\varepsilon\) is the mixture internal energy density. 
In addition, the generalized Gibbs relation/entropy balance for the mixture takes the form 
\begin{equation}
\textstyle
\dot{\eta}
=\pi+\mathop{\mathrm{div}}
\tfrac{1}{\theta}
\sum_{i}
\,\mu_{i}\bm{j}_{i}
\label{equ:RelGibGen}
\end{equation} 
and holds in the current case~\cite[e.g.,][Chapter III]{deG62}.
Here, \(\eta\) is the entropy density, \(\pi\) is the entropy production-rate density, \(\theta\) represents the absolute temperature, and \(\mu_{i}\) denotes the chemical potential of component \(i\). Note that \(\mu_{i}\) has units of energy density.  
Combining~\cref{equ:BalMasComDenMix}${}_{1}$ and~\cref{equ:RelGibGen}, one obtains the form 
\begin{equation}
\textstyle
\theta\pi
=\bm{P}\cdot\nabla\dot{\bm{\chi}} 
+\sum_{i}\mu_{i}\dot{x}_{i}
-\skew3\dot{\psi}
-\sum_{i}\bm{j}_{i}
\cdot
\nabla\mu_{i}
-\sum_{i}
\mu_{i}\sigma_{\!i}
\label{equ:DenRatDisMixIOR} 
\end{equation} 
for the mixture residual dissipation-rate density \(\theta\pi\), where \(\psi:=\varepsilon-\theta\eta\) is the mixture free energy density.

\subsection{Energetic constitutive relations} 
Due to the expected small molar fraction of hydrogen and water (parts per million range) in the solid phases, the mixture is treated for simplicity as an ideal solution with respect to~\ce{H} and~\ce{H2O}. In this case, the specific model forms 
\begin{equation}
\begin{array}{rcl}
\psi_{\mathrm{bul}}
&=&
\psi_{\mathrm{che}}
(x_{\smash{\mathrm{H}}},
x_{\smash{\mathrm{H_{2}O}}},
x_{\mathrm{O}},
\phi_{\mathrm{f}},
\phi_{\mathrm{g}},
\phi_{\mathrm{w}})
+\psi_{\mathrm{ela}}
(\nabla\bm{\chi},
\bm{F}_{\!\mathrm{R}},
\phi_{\mathrm{f}},
\phi_{\mathrm{g}},
\phi_{\mathrm{w}})
\,,\\
\psi_{\mathrm{che}}
&=&
\sum_{\alpha}
h(\phi_{\alpha})\,\psi_{\smash{\alpha}}^{\mathrm{che}}
(x_{\mathrm{O}})
+\psi_{\smash{\mathrm{H}}}(x_{\smash{\mathrm{H}}})
+\psi_{\smash{\mathrm{H_{2}O}}}(x_{\smash{\mathrm{H_{2}O}}})
\,,\\
\psi_{\mathrm{ela}}
&=&
\sum_{\alpha}
h(\phi_{\alpha})\,\psi_{\smash{\alpha}}^{\mathrm{ela}}
(\nabla\bm{\chi},\bm{F}_{\!\mathrm{R}})
\,,
\end{array}
\label{equ:DenEneFreBul}
\end{equation} 
and 
\begin{equation}
\begin{array}{rcl}
\psi_{\mathrm{int}}
&=&
\sum_{\alpha}
m(\phi_{\alpha})
\,w_{\alpha}
+\frac{1}{2}
\sum_{\alpha}\epsilon_{\alpha}
\,|\nabla\phi_{\alpha}|^{2}
+\frac{1}{2}
\,\kappa_{\mathrm{O}}
\,|\nabla x_{\mathrm{O}}|^{2}
\,,
\end{array}
\label{equ:DenEneFreInt}
\end{equation} 
are assumed for the bulk \(\psi_{\mathrm{bul}}\) and interface \(\psi_{\mathrm{int}}\) parts, respectively, of \(\psi\), i.e., 
\begin{equation}
\psi
=\psi_{\mathrm{bul}}
+\psi_{\mathrm{int}}
\,.
\label{equ:DenEneFreMix}
\end{equation} 
In \cref{equ:DenEneFreBul}, \(\psi_{\mathrm{che}}\) is the chemical part and \(\psi_{\mathrm{ela}}\) is the elastic part of \(\psi_{\mathrm{bul}}\).
The spatial presence of each phase \(\alpha\in\lbrace\hbox{f, g, w}\rbrace\) in the mixture is modeled by a corresponding non-conservative phase field \(\phi_{\alpha}\), where "f" stands for ferrite (\(\alpha\)-Fe), "g" for gas, and "w" for w\"ustite (FeO).
The order parameters $\phi_{\alpha}$ vary in a range $[0,1]$, where a value of 0 at a point in space means that the phase is occupying 0\% of the space, while a value of 1 means that it occupies 100\% of the point.
Material properties determining~\cref{equ:DenEneFreBul} and~\cref{equ:DenEneFreInt} include the energy-barrier height \(w_{\alpha}\) for the immiscibility between phases, as well as the gradient energy coefficients \(\epsilon_{\alpha}\) and \(\kappa_{\mathrm{O}}\).
In addition, \(h(x)\) and \(m(x)\) represent phase interpolation functions which will be introduced later.

The residual local deformation\footnote{The evolution of \(\bm{F}_{\!\mathrm{R}}\) is driven by stored energy reduction resulting in stress relaxation~\cite[a generalization of "stress-free strain":][]{Kha83}.}
\begin{equation}
\bm{F}_{\!\mathrm{R}}(\phi_{\mathrm{f}},\bm{F}_{\!\mathrm{P}})
=\bm{F}_{\!\mathrm{T}}(\phi_{\mathrm{f}})
\,\bm{F}_{\!\mathrm{P}}
\,,\quad
\bm{F}_{\!\mathrm{T}}
=d_{\mathrm{T}}(\phi_{\mathrm{f}})^{1/3}\bm{I}
\,,\quad
d_{\smash{\mathrm{T}}}
=1-h(\phi_{\mathrm{f}})\,\Omega_{\mathrm{f}}
\,,
\label{equ:RulFloDefLocAlt}
\end{equation} 
in the mixture is determined by that \(\bm{F}_{\!\mathrm{T}}\) due to the transformation of~\ce{FeO} into \(\alpha\)-Fe, as well as that \(\bm{F}_{\!\mathrm{P}}\) due to dislocation glide, with \(\Omega_{\mathrm{f}}\) being the relative local volume decrease during the phase transformation from FeO to \(\alpha\)-Fe.
At \(\theta=1183\) K for example, the lattice parameters of~\ce{Fe_{0.95}O} and \(\alpha\)-Fe are 0.4363 nm and 0.2907 nm, respectively \cite[][]{john1984breakdown}, resulting in a volume reduction of about 42\%~\cite[][]{mao2017reduction} and so \(\Omega_{\mathrm{f}}=0.42\). 
In the simulations to follow, \(\theta=973\) K and \(\Omega_{\mathrm{f}}=0.4\) are assumed. Whereas \(\bm{F}_{\!\mathrm{T}}\) is dilatational, note that \(\bm{F}_{\!\mathrm{P}}\) is isochoric, i.e., \(\det\bm{F}_{\!\mathrm{P}}=1\). Consequently, 
\begin{equation}
\det\bm{F}_{\!\mathrm{R}}
=(\det\bm{F}_{\!\mathrm{T}})\det\bm{F}_{\!\mathrm{P}}
=d_{\mathrm{T}}(\phi_{\mathrm{f}})
\label{equ:RulFloDefLocAltDet}
\end{equation} 
holds from \cref{equ:RulFloDefLocAlt} for the determinant of
\(\bm{F}_{\!\mathrm{R}}\). 

The phase elastic free energy density in~\cref{equ:DenEneFreBul}${}_{3}$ is given by 
\begin{equation}
\psi_{\smash{\alpha}}^{\mathrm{ela}}
=(\det\bm{F}_{\!\mathrm{R}})
\,\varphi_{\smash{\alpha}}^{\mathrm{ela}}(\bm{F}_{\!\mathrm{E}})
=d_{\mathrm{T}}(\phi_{\mathrm{f}})
\,\varphi_{\smash{\alpha}}^{\mathrm{ela}}(\bm{F}_{\!\mathrm{E}})
\,,\quad
\bm{F}_{\!\mathrm{E}}
:=(\nabla\bm{\chi})\bm{F}_{\!\smash{\mathrm{R}}}^{-1}
\,,
\label{equ:DenEneFreElaPha}
\end{equation} 
via~\cref{equ:RulFloDefLocAltDet}, where \(\varphi_{\smash{\alpha}}^{\mathrm{ela}}\) is the phase free energy per unit volume in the "intermediate" local mixture configuration, and 
\begin{equation}
\bm{F}_{\!\mathrm{E}}
:=(\nabla\bm{\chi})\bm{F}_{\!\smash{\mathrm{R}}}^{-1}
=d_{\smash{\mathrm{T}}}(\phi_{\mathrm{f}})^{-1/3}
(\nabla\bm{\chi})\bm{F}_{\!\smash{\mathrm{P}}}^{-1}
\label{equ:LocDefEla}
\end{equation}
is the elastic local deformation. Assuming isotropic elastic phase behavior with respect to this configuration, the isotropic (polyconvex) neo-Hooke form
\begin{equation}
\begin{array}{rcl}
\varphi_{\smash{\alpha}}^{\mathrm{ela}}(\bm{F}_{\!\mathrm{E}})
&=&
\tfrac{1}{4}
\,K_{\alpha}
(|\det\bm{F}_{\!\mathrm{E}}|^{2}-1-2\ln\det\bm{F}_{\!\mathrm{E}})
+\tfrac{1}{2}
\,G_{\alpha}
(|\mathop{\mathrm{uni}}\bm{F}_{\!\mathrm{E}}|^{2}-3)
\\
&=&
\tfrac{1}{4}
\,K_{\alpha}
(\det\bm{C}_{\mathrm{E}}-1-\ln\det\bm{C}_{\mathrm{E}})
+\tfrac{1}{2}
\,G_{\alpha}
(\bm{I}\cdot\mathop{\mathrm{uni}}\bm{C}_{\mathrm{E}}-3)
\\
&=&
\tfrac{1}{4}
\,K_{\alpha}
(\det\bm{B}_{\mathrm{E}}-1-\ln\det\bm{B}_{\mathrm{E}})
+\tfrac{1}{2}
\,G_{\alpha}
(\bm{I}\cdot\mathop{\mathrm{uni}}\bm{B}_{\mathrm{E}}-3)
\end{array}
\label{equ:DenEneFreIntElaPha}
\end{equation}
for \(\varphi_{\!\smash{\alpha}}^{\mathrm{ela}}\) is employed in terms of the (constant) phase bulk \(K_{\alpha}\) and shear \(G_{\alpha}\) moduli.
Here, 
\(
\mathop{\mathrm{uni}}\bm{A}
:=\bm{A}/(\det\bm{A})^{1/3}
\) 
is the unimodular part of \(\bm{A}\), 
\(
\bm{C}_{\mathrm{E}}
:=\bm{F}_{\!\smash{\mathrm{E}}}^{\mathrm{T}}
\bm{F}_{\!\smash{\mathrm{E}}}
\) 
the right, and 
\(
\bm{B}_{\mathrm{E}}
:=\bm{F}_{\!\smash{\mathrm{E}}}
\bm{F}_{\!\smash{\mathrm{E}}}^{\mathrm{T}}
\) 
the left, elastic Cauchy-Green deformation. 
Note that \cref{equ:DenEneFreElaPha} yields the reduced form 
\begin{equation}
\textstyle
\psi_{\mathrm{ela}}
=d_{\mathrm{T}}(\phi_{\mathrm{f}})
\,\varphi_{\mathrm{ela}}
(\nabla\bm{\chi},
\bm{F}_{\!\mathrm{P}},
\phi_{\mathrm{f}},
\phi_{\mathrm{g}},
\phi_{\mathrm{w}})
\,,\ \ 
\varphi_{\mathrm{ela}}
=\sum_{\alpha}
h(\phi_{\alpha})
\,\varphi_{\smash{\alpha}}^{\mathrm{ela}}
(\bm{F}_{\!\mathrm{E}}
(\nabla\bm{\chi},\phi_{\mathrm{f}},\bm{F}_{\!\mathrm{P}}))
\label{equ:DenEneFreBulEla}
\end{equation}
of \cref{equ:DenEneFreBul}${}_{3}$ for \(\psi_{\mathrm{ela}}\) which will be useful in what follows. 

Assuming the three phases (two solid phases and the gas phase) always occupy the entire mixture (i.e., no  voids, pores or cracks are in vacuum condition), the constraint 
\(
\phi_{\mathrm{f}}+\phi_{\mathrm{g}}+\phi_{\mathrm{w}}=1 
\) 
holds. Treating then 
\(\phi_{\mathrm{f}}\) and \(\phi_{\mathrm{w}}\) 
as independent, 
\begin{equation}
\phi_{\mathrm{g}}(\phi_{\mathrm{f}},\phi_{\mathrm{w}})
=1-\phi_{\mathrm{f}}-\phi_{\mathrm{w}}
\,,\quad
\skew4\dot{\phi}_{\mathrm{g}}
=-\skew4\dot{\phi}_{\mathrm{f}}
-\skew4\dot{\phi}_{\mathrm{w}}
\,,\quad
\nabla\phi_{\mathrm{g}}
=-\nabla\phi_{\mathrm{f}}-\nabla\phi_{\mathrm{w}}
\,,
\label{equ:PhaParOrdDer}
\end{equation}
follow. 
The basic constitutive assumptions and relations~\cref{equ:DenEneFreBul}-\cref{equ:PhaParOrdDer} induce the form 
\begin{equation}
\begin{array}{rcl}
\skew4\dot{\psi}
&=&
(\partial_{\smash{x_{\smash{\mathrm{H}}}}}\psi)
\,\dot{x}_{\smash{\mathrm{H}}}
+(\partial_{\smash{x_{\smash{\mathrm{H_{2}O}}}}}\psi)
\,\dot{x}_{\smash{\mathrm{H_{2}O}}}
+\partial_{\nabla\bm{\chi}}\psi\cdot\nabla\dot{\bm{\chi}}
-\bm{M}\cdot\bm{L}_{\mathrm{P}}
\\
&+&
(\delta_{\smash{x_{\smash{\mathrm{O}}}}}\psi)
\,\dot{x}_{\smash{\mathrm{O}}}
+(\delta_{\smash{\phi_{\mathrm{f}}}}\psi)
\,\skew4\dot{\phi}_{\mathrm{f}}
+(\delta_{\smash{\phi_{\mathrm{w}}}}\psi)
\,\skew4\dot{\phi}_{\mathrm{w}}
\,,
\end{array}
\label{equ:DenRatEneStoMix}
\end{equation} 
for \(\skew4\dot{\psi}\) in the mixture dissipation-rate density~\cref{equ:DenRatDisMixIOR} via the generalized no-flux boundary conditions 
\begin{equation}
\dot{x}_{\mathrm{O}}
\,\partial_{\smash{\nabla x_{\mathrm{O}}}}
\psi
\cdot
\bm{n}
=0
\,,\quad 
\skew5\dot{\phi}_{\mathrm{f}}
\,\partial_{\smash{\nabla\phi_{\mathrm{f}}}}\psi
\cdot
\bm{n}
=0
\,,\quad 
\skew5\dot{\phi}_{\mathrm{w}}
\,\partial_{\smash{\nabla\phi_{\mathrm{w}}}}\psi
\cdot
\bm{n}
=0
\,.
\label{equ:ConBouFluKin}
\end{equation} 
on the mixture boundary with outward unit normal \(\bm{n}\) relevant to purely bulk behavior.
In~\cref{equ:DenRatEneStoMix}, 
\begin{equation}
\bm{M}
:=-(\partial_{\smash{\bm{F}_{\!\mathrm{P}}}}\psi)
\bm{F}_{\!\smash{\mathrm{P}}}^{\mathrm{T}}
\label{equ:StsMan}
\end{equation}
is the Mandel stress, 
\(
\bm{L}_{\mathrm{P}}
:=\dot{\bm{F}}_{\!\mathrm{P}}\bm{F}_{\!\smash{\mathrm{P}}}^{-1}
\) 
is the inelastic "velocity gradient", and 
\(
\delta_{\!x\,}\psi
:=\partial_{\!x\,}\psi-\mathop{\mathrm{div}}\partial_{\nabla x}\psi
\) 
represents the variational derivative.
Since \(\bm{F}_{\!\mathrm{P}}\) is isochoric (unimodular), note that \(\bm{L}_{\mathrm{P}}\) is deviatoric, and 
\(
\bm{M}\cdot\bm{L}_{\mathrm{P}}
=\mathop{\mathrm{dev}}\bm{M}\cdot\bm{L}_{\mathrm{P}}
\)
holds.
Together with the dependent energetic constitutive relations 
\begin{equation}
\bm{P}=\partial_{\nabla\bm{\chi}}\psi
\,,\quad
\mu_{\mathrm{H}}
=\partial_{\smash{x_{\mathrm{H}}}}\psi
\,,\quad
\mu_{\mathrm{H_{2}O}}
=\partial_{\smash{x_{\mathrm{H_{2}O}}}}\psi
\,,\quad
\mu_{\mathrm{O}}
=\delta_{\smash{x_{\mathrm{O}}}}\psi
\,,
\label{equ:ConEneMixIOR}
\end{equation} 
for the first PK stress and component chemical potentials, respectively,~\cref{equ:DenRatEneStoMix} for \(\skew4\dot{\psi}\) results in the so-called residual form  
\begin{equation}
\begin{array}{rcl}
\theta\pi
&=&
\bm{M}\cdot\bm{L}_{\mathrm{P}}
-(\delta_{\smash{\phi_{\mathrm{f}}}}\psi)\,\skew4\dot{\phi}_{\mathrm{f}}
-(\delta_{\smash{\phi_{\mathrm{w}}}}\psi)\,\skew4\dot{\phi}_{\mathrm{w}}
-\sum_{i}\bm{j}_{i}\cdot\nabla\mu_{\mathrm{i}}
-\sum_{i}\mu_{i}\sigma_{\!i}
\end{array}
\label{equ:DenRatDisMixRedEneIOR} 
\end{equation} 
for the mixture dissipation-rate density from~\cref{equ:DenRatDisMixIOR}.

\subsection{Kinetic constitutive relations} 
For the reaction~\eqref{equ:ReaIde}, one can express \(\sigma_{\!i}\) for \(i\in\lbrace\hbox{H, H${}_{2}$O, O}\rbrace\) in the form
\begin{equation}
\sigma_{\!\mathrm{H}}
=\nu_{\smash{\mathrm{H}}}r
\,,\quad
\sigma_{\!\mathrm{H_{2}O}}
=\nu_{\smash{\mathrm{H_{2}O}}}r
\,,\quad
\sigma_{\!\mathrm{O}}
=\nu_{\smash{\mathrm{O}}}r
\,,
\label{equ:DenRatSupIOR}
\end{equation} 
~\cite[e.g.,][Chapter II]{deG62} with respect to the corresponding reaction rate \(r\), where \(\nu_{\smash{i}}\) is the true stoichiometric coefficient\footnote{Using the notation of~\cite{Pri54};~\cite{deG62} employ \(\bar{\nu}_{i}\).} of \(i\) in the reaction in \eqref{equ:ReaIde}. 
These relations reduce~\cref{equ:DenRatDisMixRedEneIOR} for the mixture dissipation-rate density to 
\begin{equation}
\begin{array}{rcl}
\theta\pi
&=&
\bm{M}\cdot\bm{L}_{\mathrm{P}}
-(\delta_{\smash{\phi_{\mathrm{f}}}}\psi)\,\skew4\dot{\phi}_{\mathrm{f}}
-(\delta_{\smash{\phi_{\mathrm{w}}}}\psi)\,\skew4\dot{\phi}_{\mathrm{w}}
\\
&-&
\bm{j}_{\mathrm{H}}\cdot\nabla\mu_{\mathrm{H}}
-\bm{j}_{\mathrm{H_{2}O}}\cdot\nabla\mu_{\mathrm{H_{2}O}}
-\bm{j}_{\mathrm{O}}\cdot\nabla\mu_{\mathrm{O}}
-\varrho r a
\,,
\end{array}
\label{equ:DenRatDisMixRedIOR} 
\end{equation} 
where 
\begin{equation}
a:=\nu_{\smash{\mathrm{H}}}\mu_{\mathrm{H}}
+\nu_{\smash{\mathrm{H_{2}O}}}\mu_{\mathrm{H_{2}O}}
+\nu_{\smash{\mathrm{O}}}\mu_{\mathrm{O}}
\label{equ:ReaForAff}
\end{equation}
is the chemical affinity \cite[e.g.,][Chapter III]{deG62} of the reaction \eqref{equ:ReaIde}. 
The reduced form~\cref{equ:DenRatDisMixRedIOR} of \(\theta\pi\) motivates in particular the kinetic (i.e., flux-force) constitutive relations 
\begin{equation}
\begin{array}{rclcrclcrcl}
\skew4\dot{\phi}_{\mathrm{f}}
&=&
-L_{\mathrm{f}}
\,\delta_{\smash{\phi_{\mathrm{f}}}}\psi
\,,&&
\skew4\dot{\phi}_{\mathrm{w}}
&=&
-L_{\mathrm{w}}
\,\delta_{\smash{\phi_{\mathrm{w}}}}\psi
\,,&&
\\ 
\bm{j}_{\mathrm{H}}
&=&
-M_{\mathrm{H}}\nabla\mu_{\mathrm{H}}
\,,&&
\bm{j}_{\mathrm{H_{2}O}}
&=&
-M_{\mathrm{H_{2}O}}\nabla\mu_{\mathrm{H_{2}O}}
\,,&&
\bm{j}_{\mathrm{O}}
&=&
-M_{\mathrm{O}}\nabla\mu_{\mathrm{O}}
\,,\\
r
&=&
-l\,a
\,,&&&&
\end{array}
\label{equ:ConKinMixIOR}
\end{equation}
in terms of the non-negative phase mobilities $L_{\mathrm{f}}$ and \(L_{\mathrm{w}}\) (units m${}^{3}$J${}^{-1}$s${}^{-1}$),  non-negative component mobilities 
\(M_{\mathrm{H}}, M_{\mathrm{H_{2}O}}, M_{\mathrm{O}}\) (units m${}^{4}$J${}^{-1}$s${}^{-1}$), and non-negative reaction kinetic coefficient \(l\) (units m${}^{3}$J${}^{-1}$s${}^{-1}$). 
In particular,~\cref{equ:ConKinMixIOR}${}_{1,2}$ represent the Ginzburg-Landau relations for overdamped non-conservative phase field dynamics. 
The effect of dislocation glide on the material behavior is modeled here for simplicity via isotropic von Mises plasticity~\cite{mises1913mechanik}, i.e., 
\begin{equation}
\bm{L}_{\mathrm{P}}
=\lambda
\,(\partial_{\bm{M}}y)
\,,\ \ 
y(\bm{M},\epsilon_{\mathrm{P}})
=|\mathop{\mathrm{dev}}\bm{M}|
-\sqrt{\tfrac{2}{3}}\ (\sigma_{\!\mathrm{Y}}+H\epsilon_{\mathrm{P}})
\leqslant 
0\,,\ \ 
\dot{\epsilon}_{\mathrm{P}}
=\lambda
\geqslant 
0
\,,\ \ 
y\lambda=0
\,,
\label{equ:PlaTwoJ}
\end{equation}
where \(\sigma_{\!\mathrm{Y}}\) is the initial yield stress, \(H\) represents the isotropic hardening modulus, and  \(\epsilon_{\mathrm{P}}\) denotes the accumulated inelastic strain. 
Since \(|\partial_{\bm{M}}y|=1\), note that \(\lambda=|\bm{L}_{\mathrm{P}}|\geqslant 0\). 
Substituting~\cref{equ:ConKinMixIOR} and~\cref{equ:PlaTwoJ} into~\cref{equ:DenRatDisMixRedIOR}, one obtains the form 
\begin{equation}
\begin{array}{rcl}
\theta\pi
&=&
\dot{\epsilon}_{\mathrm{P}}
|\mathop{\mathrm{dev}}\bm{M}|
+L_{\mathrm{f}}
|\delta_{\smash{\phi_{\mathrm{f}}}}\psi|^{2}
+L_{\mathrm{w}}
|\delta_{\smash{\phi_{\mathrm{w}}}}\psi|^{2}
\\
&+&
M_{\mathrm{H}}|\nabla\mu_{\mathrm{H}}|^{2}
+M_{\mathrm{H_{2}O}}|\nabla\mu_{\mathrm{H_{2}O}}|^{2}
+M_{\mathrm{O}}|\nabla\mu_{\mathrm{O}}|^{2}
+\varrho l|a|^{2}
\end{array}
\label{equ:DenRatDisMixIORRed} 
\end{equation} 
for the mixture dissipation-rate density which is identically non-negative and so satisfies the dissipation principle identically. 

\subsubsection{Reaction model}
Alternative to the more general flux-force relation~\cref{equ:ConKinMixIOR}${}_{6}$ for the reaction rate \(r\), one can also work with a reaction model. 
Perhaps the simplest such model is represented by the law of mass action~\cite[e.g.,][Chapter X]{deG62}.
For the reaction \eqref{equ:ReaIde}, this takes the form 
\(
r=\kappa_{\smash{\mathrm{for}}}
x_{\smash{\mathrm{O}}}^{-\nu_{\smash{\mathrm{O}}}}
x_{\smash{\mathrm{H}}}^{-\nu_{\smash{\mathrm{H}}}}
-\kappa_{\smash{\mathrm{rev}}}
x_{\smash{\mathrm{H_{2}O}}}^{\nu_{\smash{\mathrm{H_{2}O}}}}
\),
with 
\(
\nu_{\smash{\mathrm{H}}}
=-2
\),
\(
\nu_{\smash{\mathrm{O}}}
=-1
\), 
and 
\(
\nu_{\smash{\mathrm{H_{2}O}}}
=1
\). 
Here, \(\kappa_{\mathrm{for}}\) is the rate coefficient of the forward, and \(\kappa_{\mathrm{rev}}\) that of the reverse, reaction. 
In particular, this is relevant to ideal homogeneous reaction cases where the "solvent" (in the current case, Fe) is not part of the reaction. More generally, a number of empirical relations for \(r\) \cite[e.g.,][]{newman2012electrochemical,bazant2013theory} deviate from the law of mass action.
Experimental results for~\eqref{equ:ReaIde}~\cite{Bai2018,Liu2014} are consistent with the empirical form 
\begin{equation}
r=\kappa_{\smash{\mathrm{for}}}
x_{\smash{\mathrm{O}}}x_{\smash{\mathrm{H}}}
\label{equ:EmpModRea}
\end{equation}
for \(r\) with \(\kappa_{\smash{\mathrm{rev}}}\approx 0\). Since the reaction~\eqref{equ:ReaIde} takes place only when the gas phase is present, 
\begin{equation}
\label{eq:forward-k}
\kappa_{\smash{\mathrm{for}}}
=m(\phi_{\mathrm{g}})\,k_{\smash{\mathrm{for}}}
\end{equation}
is assumed here, with \(k_{\smash{\mathrm{for}}}\) constant, and \(m(x)\) the interpolation function in~\cref{equ:DenEneFreInt}.

\subsection{Summary of derived model relations}
The current model formulation yields in particular the field relations 
\begin{equation}
\begin{array}{rcl}
\dot{x}_{\mathrm{H}}
&=&
\mathop{\mathrm{div}}
D_{\mathrm{H}}
\nabla x_{\mathrm{H}}
-2m(\phi_{\mathrm{g}})\,k_{\smash{\mathrm{for}}}
x_{\smash{\mathrm{O}}}x_{\smash{\mathrm{H}}}
\,,\\
\dot{x}_{\mathrm{H_{2}O}}
&=&
\mathop{\mathrm{div}}
D_{\mathrm{H_{2}O}}
\nabla x_{\mathrm{H_{2}O}}
+m(\phi_{\mathrm{g}})\,k_{\smash{\mathrm{for}}}
x_{\smash{\mathrm{O}}}x_{\smash{\mathrm{H}}}
\,,\\ 
\dot{x}_{\mathrm{O}}
&=&
\mathop{\mathrm{div}}
\lbrack
D_{\mathrm{O}}\nabla x_{\mathrm{O}}
-M_{\mathrm{O}}
\mathop{\mathrm{div}}\nabla(\kappa_{\mathrm{O}}\nabla x_{\mathrm{O}})
\rbrack
-m(\phi_{\mathrm{g}})\,k_{\smash{\mathrm{for}}}
x_{\smash{\mathrm{O}}}x_{\smash{\mathrm{H}}}
\,,\\
\bm{0}
&=&
\mathop{\mathrm{div}}
\partial_{\nabla\bm{\chi}}\psi
\,,\\
\skew4\dot{\phi}_{\mathrm{f}}
&=&
L_{\mathrm{f}}
\mathop{\mathrm{div}}\partial_{\smash{\nabla\phi_{\mathrm{f}}}}\psi
-L_{\mathrm{f}}
\,\partial_{\smash{\phi_{\mathrm{f}}}}\psi
\,,\\
\skew4\dot{\phi}_{\mathrm{w}}
&=&
L_{\mathrm{w}}
\mathop{\mathrm{div}}\partial_{\smash{\nabla\phi_{\mathrm{w}}}}\psi
-L_{\mathrm{w}}
\,\partial_{\smash{\phi_{\mathrm{w}}}}\psi
\,,
\end{array}
\label{equ:FieRelDRI}
\end{equation} 
for the unknown fields \(x_{\ce{H}}\), \(x_{\ce{H2O}}\), \(x_{\ce{O}}\), 
\(\bm{\chi}\), 
\(\phi_{\mathrm{f}},\phi_{\mathrm{w}}\) with 
\begin{equation}
D_{i}
:=M_{i}
\,(\partial_{\smash{x_{i}}}^{\,2}\psi)
\label{equ:DifCom}
\end{equation}
the component molar-fraction-based diffusivity. 
In turn, \(\phi_{\mathrm{f}}\) and \(\phi_{\mathrm{w}}\) determine \(\phi_{\mathrm{g}}\) via~\cref{equ:PhaParOrdDer}. 
In particular,~\cref{equ:FieRelDRI}${}_{1-3}$ represent reduced forms of component mass balance~\cref{equ:BalMasComDenMix}${}_{1}$ via\\~\cref{equ:ConEneMixIOR}${}_{2-4}$,~\cref{equ:DenRatSupIOR},~\cref{equ:ConKinMixIOR}${}_{3,4}$, and~\cref{equ:EmpModRea}. 
Likewise,~\cref{equ:FieRelDRI}${}_{4}$ is obtained from the mixture linear momentum balance~\cref{equ:BalMasComDenMix}${}_{2}$ via~\cref{equ:ConEneMixIOR}${}_{1}$, and~\cref{equ:FieRelDRI}${}_{5,6}$ follow from the Ginzburg-Landau relations~\cref{equ:ConKinMixIOR}${}_{1,2}$. 
It should be mentioned that the molecular size of water is substantially greater than that of mono-atomic  hydrogen and oxygen, therefore, the diffusivity of water in the solid phase (w\"ustite and $\alpha$-iron) is tuned to a negligible value.

The free energy model relations~\cref{equ:DenEneFreBul}-~\cref{equ:DenEneFreMix},~\cref{equ:DenEneFreElaPha},~\cref{equ:DenEneFreIntElaPha} and~\cref{equ:DenEneFreBulEla} determine the model forms
\begin{equation}
\begin{array}{rcl}
\partial_{\nabla\bm{\chi}}\psi
&=&
K\,(\det\bm{B}_{\mathrm{E}}-1)
\,(\nabla\bm{\chi})^{-\mathrm{T}}
+2\,G\,(\mathop{\mathrm{dev}}
\mathop{\mathrm{uni}}
\bm{B}_{\mathrm{E}})
\,(\nabla\bm{\chi})^{-\mathrm{T}}
\,,\\
\bm{M}
&=&
K\,(\det\bm{C}_{\mathrm{E}}-1)\,\bm{I}
+2\,G\mathop{\mathrm{dev}}
\mathop{\mathrm{uni}}
\bm{C}_{\mathrm{E}}
\,,
\end{array}
\label{equ:StsRelMod}
\end{equation} 
for the first PK and Mandel stresses, respectively, via~\cref{equ:LocDefEla},~\cref{equ:StsMan}, and~\cref{equ:ConEneMixIOR}${}_{1}$, in the context of finite elastic strain, with 
\begin{equation}
\textstyle
K:=d_{\mathrm{T}}(\phi_{\mathrm{f}})
\sum_{\alpha}h(\phi_{\alpha})\,K_{\alpha}
\,,\quad
G:=d_{\mathrm{T}}(\phi_{\mathrm{f}})
\sum_{\alpha}h(\phi_{\alpha})\,G_{\alpha}
\,,
\label{equ:ModElaEff}
\end{equation}
the effective mixture elastic moduli. 
Likewise, 
\begin{equation}
\begin{array}{rcl}
\partial_{\smash{\phi_{\mathrm{f}}}}\psi
&=&
h^{\prime}(\phi_{\mathrm{f}})
\,(\psi_{\smash{\mathrm{f}}}^{\mathrm{che}}
+\psi_{\smash{\mathrm{f}}}^{\mathrm{ela}})
-h^{\prime}(\phi_{\mathrm{g}})
\,(\psi_{\smash{\mathrm{g}}}^{\mathrm{che}}
+\psi_{\smash{\mathrm{g}}}^{\mathrm{ela}})
+m^{\prime}(\phi_{\mathrm{f}})
\,w_{\smash{\mathrm{f}}}
-m^{\prime}(\phi_{\mathrm{g}})
\,w_{\smash{\mathrm{g}}}
\\
&+&
d_{\smash{\mathrm{T}}}^{\prime}\,\varphi_{\mathrm{ela}}
-\tfrac{1}{3}\,\bm{I}\cdot\bm{M}
\,,\\
\partial_{\smash{\phi_{\mathrm{w}}}}\psi
&=&
h^{\prime}(\phi_{\mathrm{w}})
\,(\psi_{\smash{\mathrm{w}}}^{\mathrm{che}}
+\psi_{\smash{\mathrm{w}}}^{\mathrm{ela}})
-h^{\prime}(\phi_{\mathrm{g}})
\,(\psi_{\smash{\mathrm{g}}}^{\mathrm{che}}
+\psi_{\smash{\mathrm{g}}}^{\mathrm{ela}})
+m^{\prime}(\phi_{\mathrm{w}})
\,w_{\smash{\mathrm{w}}}
-m^{\prime}(\phi_{\mathrm{g}})
\,w_{\smash{\mathrm{g}}}
\,,\\
\partial_{\smash{\nabla\phi_{\mathrm{f}}}}\psi
&=&
\epsilon_{\mathrm{f}}\nabla\phi_{\mathrm{f}}
-\epsilon_{\mathrm{g}}\nabla\phi_{\mathrm{g}}
\,,\\
\partial_{\smash{\nabla\phi_{\mathrm{w}}}}\psi
&=&
\epsilon_{\mathrm{w}}\nabla\phi_{\mathrm{w}}
-\epsilon_{\mathrm{g}}\nabla\phi_{\mathrm{g}}
\,,
\label{equ:FiePhaDenEneFreDerPar}
\end{array}
\end{equation} 
are determined by the current free energy model via~\cref{equ:DenEneFreBul},~\cref{equ:DenEneFreInt},~\cref{equ:RulFloDefLocAlt},~\cref{equ:PhaParOrdDer} and~\cref{equ:StsMan}.
In the first of these \(\partial_{\smash{\phi_{\mathrm{f}}}}\psi\) appears the spherical part
\(
\tfrac{1}{3}\,\bm{I}\cdot\bm{M}=K\,(\det\bm{C}_{\mathrm{E}}-1)
\) 
of \(\bm{M}\) from~\cref{equ:StsRelMod}${}_{2}$.
The evolution relation 
\begin{equation}
\dot{\bm{F}}_{\!\mathrm{P}}
=\bm{L}_{\mathrm{P}}
\bm{F}_{\!\smash{\mathrm{P}}}
\,,\quad
\bm{L}_{\mathrm{P}}
=\lambda
\,\mathop{\mathrm{dev}}\bm{M}/|\mathop{\mathrm{dev}}\bm{M}|
\,,
\label{equ:RulFloPla}
\end{equation}
for \(\bm{F}_{\!\smash{\mathrm{P}}}\) from \cref{equ:PlaTwoJ} is determined by the deviatoric part 
\(
\mathop{\mathrm{dev}}\bm{M}
=2\,G\mathop{\mathrm{dev}}
\mathop{\mathrm{uni}}
\bm{C}_{\mathrm{E}}
\)
of \(\bm{M}\) from~\cref{equ:StsRelMod}${}_{2}$.
Together with \(\nabla\bm{\chi}\) and \(\phi_{\mathrm{f}}\), \(\bm{F}_{\!\smash{\mathrm{P}}}\) determines the elastic local deformation \(\bm{F}_{\!\mathrm{E}}\) via~\cref{equ:LocDefEla}. 
In turn, \(\bm{F}_{\!\mathrm{E}}\), \(\phi_{\mathrm{f}}\) and \(\phi_{\mathrm{w}}\) determine \(\bm{P}\), \(\bm{M}\) and \(\varphi_{\mathrm{ela}}\), the latter appearing in~\cref{equ:FiePhaDenEneFreDerPar}${}_{1}$.

The (weak form of the) fields and auxiliary relations~\cref{equ:FieRelDRI}-\cref{equ:RulFloPla} have been implemented in the open source finite element package libMesh/MOOSE~\cite[][]{kirk2006libmesh,gaston2009moose}. 
The backward euler (BE) method has been used for the time integration of~\cref{equ:FieRelDRI}${}_{1-3}$ and \cref{equ:FieRelDRI}${}_{5-6}$, and also the plastic flow in~\cref{equ:RulFloPla}.
These fully coupled equations are solved together with the help of routines from PETSc \cite[][]{balay2019petsc} and an iterative solver based on the Preconditioned Jacobian-free Newton Krylov (PJFNK) method implemented in MOOSE. 

\section{Model identification}
\label{sec:calib} 
\subsection{Determination of chemical energy using CALPHAD}
\begin{figure}[H]
	\centering
	\includegraphics[width=0.6\linewidth]{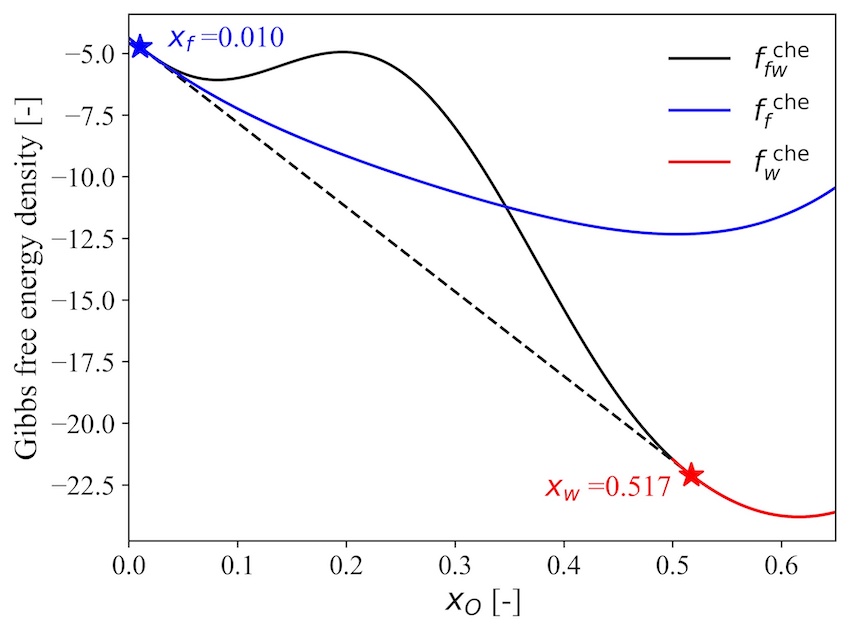}
	\caption{Results for 
		\(
		f_{\smash{\mathrm{f}}}^{\mathrm{che}}(x_{\mathrm{O}})
		\) 
		(blue curve),
		\(
		f_{\smash{\mathrm{w}}}^{\mathrm{che}}(x_{\mathrm{O}})
		\)
		(red curve), and 
		\(
		f_{\smash{\mathrm{fw}}}^{\mathrm{che}}(x_{\mathrm{O}})
		\)
		(black curve) 
		at \(\theta=973.15\) K and room pressure based on the Thermo-Calc TCOX10 database~\cite{andersson2002thermo,sundman1991assessment,hidayat2015thermodynamic}. The blue star marks the equilibrium oxygen mole fraction 
		$x_{\mathrm{f}}=0.01$ in ferrite, and the red star that 
		$x_{\mathrm{w}}=0.571$ in w\"ustite.
	}
	\label{fig:free-energy-fitting}
\end{figure}

In HyDRI, the transformation from w\"ustite to ferrite is observed to occur at temperatures above 843 K~\cite{KIM2021116933}. In this work, \(\theta=973.15\) K is assumed for the simulations below, the Thermo-Calc TCOX10 database~\cite{andersson2002thermo,sundman1991assessment,hidayat2015thermodynamic} is used to determine the forms 
\begin{equation}
\begin{array}{rcl}
f_{\smash{\mathrm{f}}}^{\mathrm{che}}(x_{\smash{\ce{O}}})
&=&
163.708\,x_{\smash{\ce{O}}}^4
-182.510\,x_{\smash{\ce{O}}}^3
+90.573\,x_{\smash{\ce{O}}}^2
-36.031\,x_{\smash{\ce{O}}}
-4.372
\,,\\
f_{\smash{\mathrm{w}}}^{\mathrm{che}}(x_{\smash{\ce{O}}})
&=&
173.263\,x_{\smash{\ce{O}}}^2
-213.460\,x_{\smash{\ce{O}}}
+41.953
\,,
\end{array}
\label{equ:DenEneFreChePha}
\end{equation} 
of 
\(
f_{\smash{\alpha}}^{\mathrm{che}}(x_{\smash{\ce{O}}})
:=(V_{\mathrm{mol}}/R\theta)
\,\psi_{\smash{\alpha}}^{\mathrm{che}}(x_{\smash{\ce{O}}})
\) 
for ferrite and w\"ustite, respectively, in~\cref{equ:DenEneFreBul}${}_{2}$, 
where \(R\) is the gas constant, and \(V_{\mathrm{mol}}\) the molar volume of w\"ustite. Since \(x_{\smash{\ce{O}}}\) in the gas phase 
(consisting mainly of hydrogen and water) is nearly zero, the dilute form 
\begin{equation}
f_{\smash{\mathrm{g}}}^{\mathrm{che}}(x_{\smash{\ce{O}}})
=10^{3}\,x_{\smash{\ce{O}}}^{2}
\label{eq:free-energy-gas-phase}
\end{equation}
for the scaled chemical free energy density of the gas phase is adopted here, where $10^3$ has been used to constrain oxygen content to be zero in the gas phase.
Based on~\cref{equ:DenEneFreChePha} and the choice \(h(x)=x^{3}(6x^{2}-15x+10)\), 
\begin{equation}
f_{\smash{\mathrm{fw}}}^{\mathrm{che}}(x_{\mathrm{O}})
:=h(\phi_{\mathrm{f}})\,f_{\smash{\mathrm{f}}}^{\mathrm{che}}
(x_{\mathrm{O}})
+h(\phi_{\mathrm{w}})\,f_{\smash{\mathrm{w}}}^{\mathrm{che}}
(x_{\mathrm{O}})
\,,\ \ 
\phi_{\mathrm{f}}=\frac{x_{\mathrm{w}}-x_{\ce{O}}}{x_{\mathrm{w}}-x_{\mathrm{f}}}
\,,\ \ 
\phi_{\mathrm{w}}=\frac{x_{\ce{O}}-x_{\mathrm{f}}}{x_{\mathrm{w}}-x_{\mathrm{f}}}
\,,
\label{equ:ScaEneFreCheWusFer}
\end{equation} 
is determined for \(x_{\mathrm{f}}\leqslant x_{\ce{O}}\leqslant x_{\mathrm{w}}\) via the common tangent construction~\cite[][]{reichl1999modern,cengel2007thermodynamics}. The results~\eqref{equ:DenEneFreChePha} and~\eqref{equ:ScaEneFreCheWusFer} are displayed in~\cref{fig:free-energy-fitting}. 
\noindent The actual oxygen content in ferrite is in fact lower (0.008 pct:~\cite{seybolt1954solubility}) than $x_{\mathrm{f}}=0.01$. For simplicity, however, this value for \(x_{\mathrm{f}}\) is assumed for the numerical simulations. 

\subsection{Constant parameter values}
Values for the constant material properties and model parameters employed 
in the simulations are listed in Table~\ref{Tab_matProp}.
\begin{table}[H]
	\centering
	\begin{tabular}{|l|c|c|c|l|} \hline
		Quantity & Symbol & Value & Units 
		\\ \hline 
		Molar volume & $V_{\mathrm{mol}}$ 
		& $1.224{}\times 10{}^{-5}$~\cite{robie1962molar} & m${}^{3}$ mol${}^{-1}$ 
		\\ 
		Phase energy barrier (f,g,w) & $w_{\alpha}$ 
		& 1.0${}\times 10{}^{3}\,V_{\smash{\mathrm{mol}}}^{-1}$~$^{*}$ & J m${}^{-3}$ 
		\\ 
		Phase gradient energy (f,g,w) & $\varepsilon_{\alpha}$ 
		& $4.0{}\times10{}^{-5}$~$^{*}$ & J m${}^{-1}$ 
		\\ 
		Chemical gradient energy & $\kappa_{\mathrm{O}}$ 
		& $5.0{}\times10{}^{-10}$~$^{*}$ & J m${}^{-1}$ 
		\\ 
		Relative volume change & \(\Omega_{\mathrm{f}}\) 
		& 0.4~\cite{mao2017reduction} & -
		\\ 
		Bulk and shear moduli (f) & \(K_{\mathrm{f}}, G_{\mathrm{f}}\) 
		& 85, 41~\cite{sumino1980elastic} & GPa 
		\\
		Bulk and shear moduli (g) & \(K_{\mathrm{g}}, G_{\mathrm{g}}\) 
		& 0.17, 0.082 $^{*}$ & GPa 
		\\
		Bulk and shear moduli (w) & \(K_{\mathrm{w}}, G_{\mathrm{w}}\) 
		& 88, 24~\cite{sumino1980elastic} & GPa 
		\\
		Yield stress, hardening modulus & \(\sigma_{\mathrm{Y}}, H\) 
		& 0.3, 1.29~\cite{wang2013effect,wang2018mechanical} & GPa 
		\\ 
		Oxygen diffusivity (f,w) & \(D_{\mathrm{O}}\) 
		& $2.16{}\times10{}^{-11}$~\cite{mao2017reduction} & m${}^{2}$ s${}^{-1}$ 
		\\ 
		Oxygen diffusivity (g) & \(D_{\mathrm{O}}\) 
		& $1.0{}\times10{}^{-8}$~$^{*}$ & m${}^{2}$ s${}^{-1}$ 
		\\ 
		Hydrogen diffusivity (f,w) & \(D_{\mathrm{H}}\) 
		& 2.5${}\times10{}^{-11}$~\cite{sojka2016diffusion} & m${}^{2}$ s${}^{-1}$ 
		\\ 
		Hydrogen diffusivity (g) & \(D_{\mathrm{H}}\) 
		& 1.0${}\times10{}^{-8}$~$^{*}$ & m${}^{2}$ s${}^{-1}$ 
		\\ 
		Water diffusivity (f,w) & \(D_{\ce{H2O}}\) 
		& 2.5${}\times10{}^{-20}$~$^{*}$ & m${}^{2}$ s${}^{-1}$ 
		\\ 
		Water diffusivity (g) & \(D_{\ce{H2O}}\) 
		& 1.0${}\times10{}^{-8}$~$^{*}$ & m${}^{2}$ s${}^{-1}$ 
		\\ 
		Phase mobility (f,g,w) & $L_{\alpha}$ 
		& 1.5${}\times10{}^{-10}$~$^{*}$ & J${}^{-1}$ m${}^{3}$ s${}^{-1}$ 
		\\ 
		Reaction constant & \(k_{\mathrm{for}}\) 
		& 4.5${}\times10^{4}$~\cite{Bai2018,Liu2014} & s${}^{-1}$ 
		\\ \hline
	\end{tabular}
	\caption{Material properties assumed for HyDRI at \(\theta=973.15\) K and room pressure. Phases are indicated in parentheses. All values in SI units. The superscript \(\ast\) denotes a parameter used in this work. See text for details.}
	\label{Tab_matProp}
\end{table}
\noindent 

Due to the lack of experimental data and reference values, the parameters in~\cref{Tab_matProp} with the superscript \(\ast\) have been used in this work.
For instance, the phase gradient energy $\varepsilon_{\alpha}=4.0\times 10^{-5}~\si{\joule\meter^{-1}}$ and the chemical gradient energy $\kappa_{\mathrm{O}}=5.0\times10^{-10}~\si{\joule\meter^{-1}}$ have been used to get a well-resolved interfaces in the simulation.
To achieve a reasonable phase transformation rate, the mobility $L_{\alpha}=1.5\times 10^{-10}~\si{\joule^{-1}\meter^{3}\second^{-1}}$ for each phase has been utilized.
Since water can not diffuse into the solid phase, $D_{\ce{H2O}}=2.5\times10^{-20}$ is employed for w\"ustite and $\alpha$-iron in the simulation.
Considering that, the diffusion of oxygen, hydrogen, and water in the gas phase is much faster than in the solid phase, $D_{\ce{O}}=D_{\ce{O}}=D_{\ce{H2O}}=10^{-8}~\si{\meter^{2}\second^{-1}}$ have been used.
The value of \(D_{\mathrm{H}}\) for the solid phase is taken from~\cite{sojka2016diffusion}. The value of \(D_{\mathrm{O}}\) for solid phases is based on 
\(
D_{\mathrm{O}}
=3.7\times 10^{-7}e^{-Q/R\theta}
\),
with \(Q=98000\) J/mol \cite[][]{mao2017reduction}, again at \(\theta=973.15\) K. 
Whereas \(D_{\mathrm{O}}\) is assumed constant here, note that 
\(
M_{\mathrm{O}}
=D_{\mathrm{O}}
\,(\partial_{\smash{x_{\mathrm{O}}}}^{\,2}\psi_{\mathrm{che}})^{-1}
\) 
depends on \(x_{\mathrm{O}}\) through \(\psi_{\mathrm{che}}\).
To get rid of the numerical singularity, the bulk modulus $K_{g}$ and shear modulus $G_{g}$ of the gas phase are set to be a small value, were $K_{g}=0.17~\si{\giga\pascal}$ and $G_{g}=0.082~\si{\giga\pascal}$ are used, respectively.

\section{Results and discussion}
\label{sec:results}

\subsection{Benchmark cases}
\label{sec:benchmark}
\begin{figure}[H]
\centering
\includegraphics[width=0.99\linewidth]{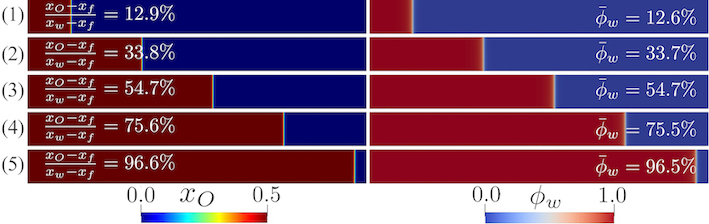}
\caption{W\"ustite (red) and $\alpha$-iron (blue) phase equilibrium study at different oxygen molar fraction values of 0.1, 0.2, 0.3, 0.4 and 0.5, from (1) to (5), respectively. The left column shows the oxygen molar fraction and the right column shows the w\"ustite order parameter.}
\label{fig:oxygen-dependency}
\end{figure}

A series of phase equilibrium simulations are performed to study the phase fractions under different oxygen content. We neglect the chemical reaction and mechanical coupling when determining the oxygen dependence of free energies in this case.
A rectangular domain with a size of~\si{5~\micro\meter}$\times$~\si{1~\micro\meter} is used for the simulations. We set the initial oxygen molar fraction homogeneously constant across the entire simulation domain, but only half of the rectangle is occupied by the $\alpha$-iron phase, while the other half is the w\"ustite phase.
For different initial oxygen contents, the phase evolution and equilibrium oxygen content contour plots are shown in~\cref{fig:oxygen-dependency}. The equilibrium results for different initial $x_\mathrm{O}$ values of 0.1, 0.2, 0.3, 0.4, and 0.5 are shown in~\cref{fig:oxygen-dependency}(1)-(5), respectively. Using the~\textit{lever rule}~\cite{william2006foundations}, we analytically find the phase fraction as $\frac{x_\mathrm{O}-x_{f}}{x_{w}-x_{f}}$. As a comparison, the numerical phase fraction $\bar{\phi}_{w}$ is determined by integrating the w\"ustite phase order parameter $\phi_{w}$ over the entire domain at equilibrium state as
\begin{equation}
\label{eq:w-bar}
\bar{\phi}_{w}=\frac{\int_{\Omega}\phi_{w}dV}{\int_{\Omega} dV}.
\end{equation}
As seen from~\cref{fig:oxygen-dependency}, the phase fraction results based on the simulation agree well with the analytical calculations for different initial oxygen contents. This serves as a first validation of the implementation as well as the energy model applied to a simple test scenario that is accessible to an analytical solution. 

\subsection{Iron ore reduction in an extended core-shell model}
In this section, the reduction reaction in a core-shell model is investigated. Analytical versions of core-shell models are often used to investigate solid-gas reactions without considering the effects of microstructure or mechanics.
Here we employ the developed chemo-mechanically coupled PF model to a simple core-shell geometry.
The iron oxide sample is contained in a 4.2~\si{\micro\meter}-diameter core embedded in a 5~\si{\micro\meter} gas-filled ($\phi_{g}=1$) simulation box.
Initially, for the solid phase, we set $\phi_{w}=1$, $\phi_{f}=\phi_{g}=0$ and the oxygen content to the equilibrium oxygen content in w\"ustite, i.e. $x_{w}$. For the gas phase, $x_{O}=0$ is assumed. Due to the symmetry of the sample, only 1/4 of the whole sample is considered in this simulation.
It is worth noting that the gas phase is initially assumed to only consist of hydrogen. In the solid phase, however, the hydrogen level is exceedingly low (10$\sim$40 ppm). As a consequence, we set the maximum hydrogen level in the gas phase to a value of 1\% in order to attain a reasonable result.

\subsubsection{Reduction reaction in an elastic system}
\label{sec:stress-influence}
\begin{figure}[H]
\centering
\includegraphics[width=1.02\linewidth]{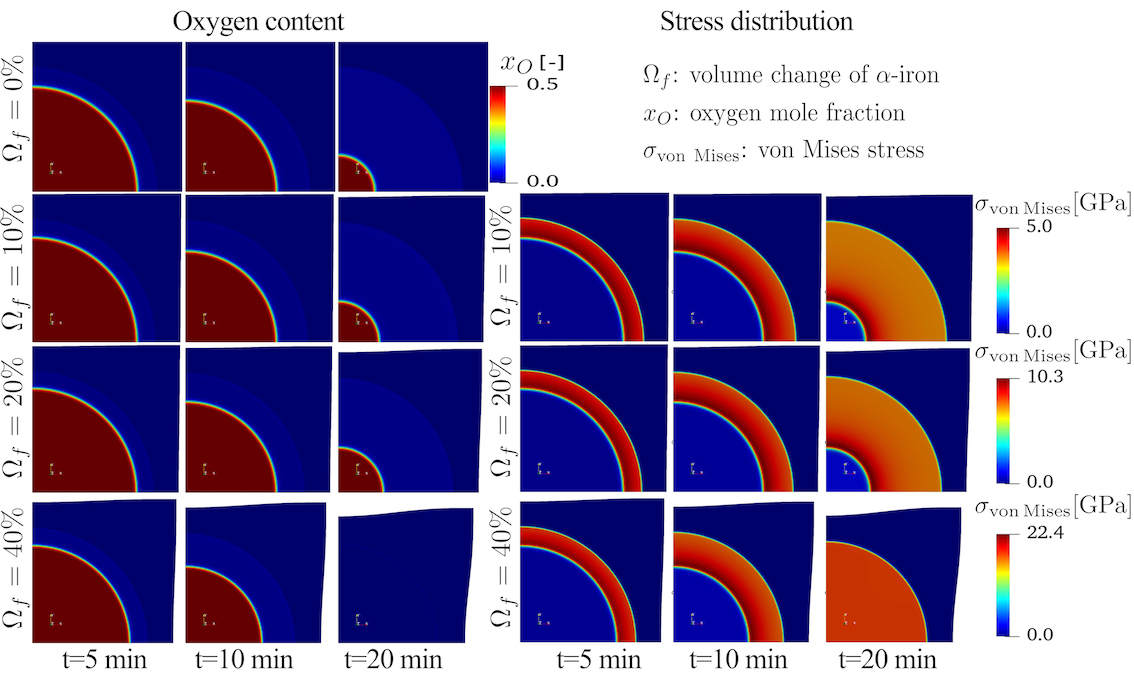}
\caption{Distribution of the oxygen content (left) and von Mises stress (right) at different simulation times for several transformation volume change of $\alpha$-iron.}
\label{fig:coreshell-elastic-result}
\end{figure}

The PF model can accurately predict the oxygen molar fraction distribution as well as the w\"ustite phase evolution over time, as shown in~\cref{sec:benchmark}.
Therefore, the fully coupled model presented in~\cref{sec:model} is used in this section to investigate the influence of mechanical stresses on the reduction process.
The volume change of w\"ustite and gas phase is set to zero. To qualitatively investigate the effects of mechanical stresses, several $\Omega_{f}$ values are considered in the simulation sets shown in~\cref{fig:coreshell-elastic-result}.
At the reaction time t=5~\si{\minute}, the different volume change induced by the phase transformation results in a similar transition region between $x_{f}$ (the outer regions of the core) and $x_{w}$ (the center of the core), as shown in the figures in the first column of~\cref{fig:coreshell-elastic-result} (left).
This is attributed to the low stress level at the beginning of the reduction reaction where not yet so much of the oxygen has been lost due to the redox reaction.
As the reduction proceeds, stress builds up, and the transition area in the sample with the higher volume change of $\alpha$-iron phase ($\Omega_{f}$) becomes even larger (see figures in the middle-bottom and the right-bottom).
Finally, the sample with the highest $\Omega_{f}$ value (40\%) has the lowest oxygen content thus achieving the highest reduction degree among all the cases.
Accordingly, the von Mises stress distribution is plotted in~\cref{fig:coreshell-elastic-result} (right). We find that a higher volume change results in higher stress levels as expected.
For instance, for a sample with a 40\% volume change, the stress could reach 22.4~\si{\giga\pascal}, while the stress level is 5.0~\si{\giga\pascal} in the case of a 10\% volume change.
Note that this simple example scenario considers only purely elastic material behaviour. The effect of elasto-plastic response on the reduction is investigated in the next subsection. 
The results show that the maximum stress is located at the w\"ustite--$\alpha$-iron phase interface.
This is also expected, due to the high mechanical contrast and volume change across this interface associated with the structural phase transformation and the oxygen loss.

\begin{figure}[H]
	\centering
	\includegraphics[width=0.995\linewidth]{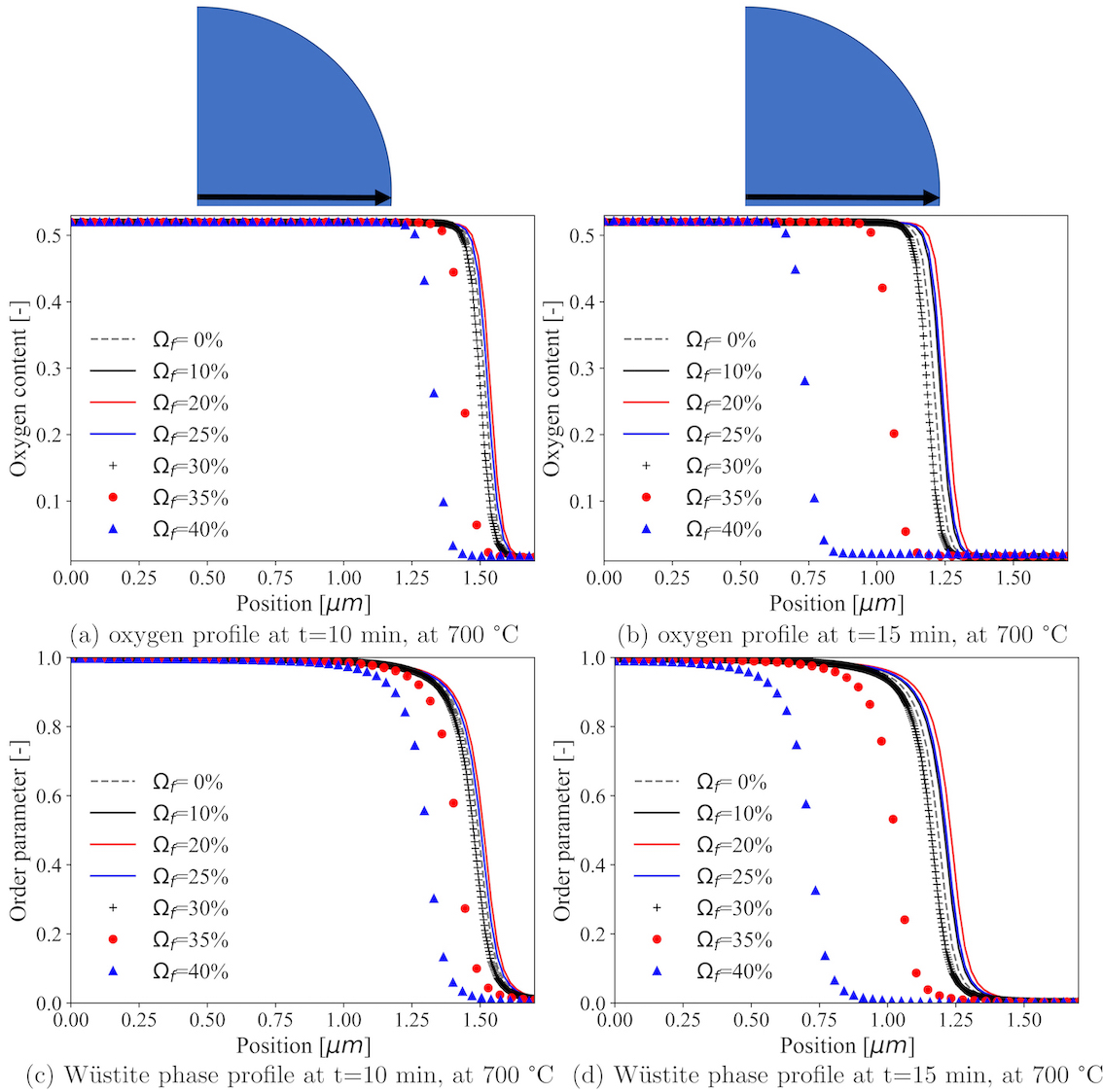}
	\caption{Spatial distribution of the oxygen molar fraction and w\"ustite phase order parameter under different volume change. The center of the core is located at $x=0.0~\si{\micro\meter}$, while the surface of the core is set at $x=1.7~\si{\micro\meter}$ (the inner side of the core's surface). (a) and (b) show the oxygen profile along the radius of the core at 10 and 15~\si{\minute}, respectively. (c) and (d) show the w\"ustite phase order parameter value along the radius of the core at 10 and 15~\si{\minute}, respectively. The dashed/solid lines with different colors indicate the cases with small volume change ($\Omega_{f}\leq 25\%$), while the markers represent the cases with large volume change ($\Omega_{f}\geq 30\%$).}
	\label{fig:coreshell-profiles}
\end{figure}

To further investigate the influence of mechanics on the w\"ustite reduction, the oxygen content and w\"ustite phase order parameter at different simulation times are plotted along the radius of the core-shell model, as shown in~\cref{fig:coreshell-profiles}. 
Based on the simulation results shown in~\cref{fig:coreshell-profiles}, the increase of the transformation volume change ($\Omega_{f}$) results in a non-monotonic change in the reduction kinetics.
From the volume change of $\alpha$-iron from 0\% to 25\%, increasing $\Omega_{f}$ causes a slight drop in the reaction kinetics (the red and black curves are behind and close to the dashed curve in~\cref{fig:coreshell-profiles}(b) for a reduction duration of t=15~\si{\minute} for example).
In contrast to this regime, raising the volume change of $\alpha$-iron from $\Omega_{f}=30\%$ to $\Omega_{f}=40\%$ promotes faster reaction kinetics.
This change in reaction kinetics is attributed to the effect of the elastic free energy on the system.
At the lower end of the volume change ($\Omega_{f}\leq 25\%$), the elastic free energy poses a substantial extra energy cost associated with the transformation from w\"ustite to  $\alpha$-iron.
As the volume change of the $\alpha$-iron phase increases, the transformation induced stresses and thus also the elastic energy increases.
This leads to a larger energy penalty for the transformation, slowing down the overall reduction kinetics.
At the higher end of the volume change of the $\alpha$-iron phase ($\Omega_{f}> 25\%$), the induced stresses and the elastic energy are considerable and they significantly influence the shape of the total free energy of the system.
This aspect is examined more closely at the end of this subsection.

\begin{figure}[H]
	\centering
	\begin{subfigure}{.495\textwidth}
		\centering
		\includegraphics[width=1.02\linewidth]{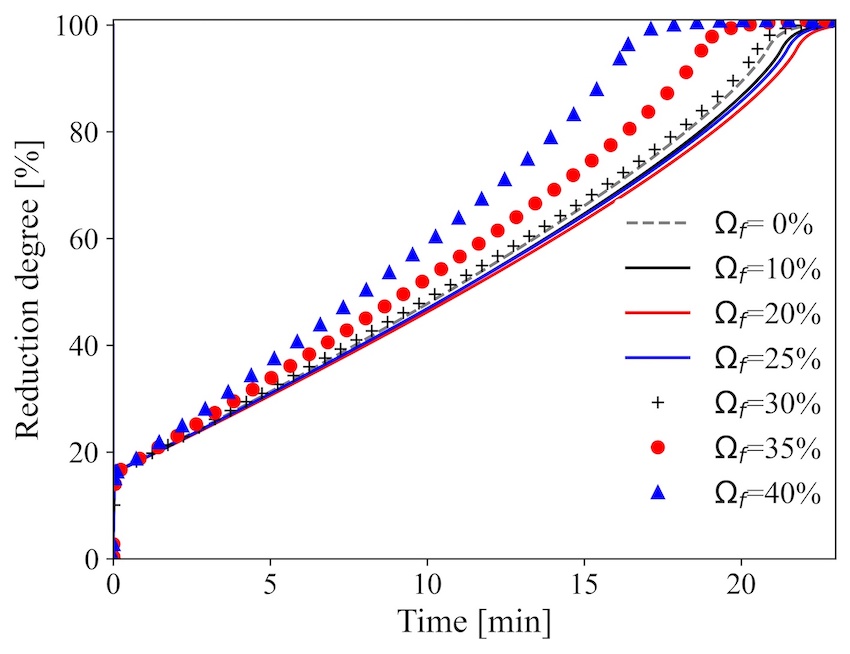}
		\caption{reduction degree at 700~\si{\degreeCelsius}}
		\label{fig:coreshell-reduction-degree}
	\end{subfigure}
	\begin{subfigure}{.495\textwidth}
		\centering
		\includegraphics[width=1.02\linewidth]{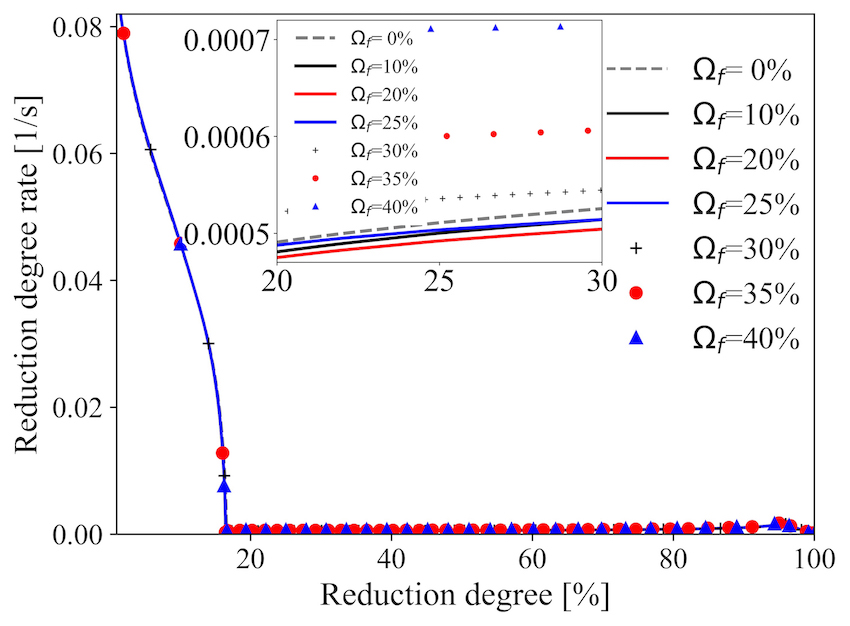}
		\caption{reduction degree rate at 700~\si{\degreeCelsius}}
		\label{fig:coreshell-reduction-degree-rate}
	\end{subfigure}
	\caption{Profiles of (a) reduction degree, and (b) reduction degree rate. Solid lines of varying colors represent samples with a small $\Omega_{f}$ value ($\leq$25\%). Markers of various colors are employed for large $\Omega_{f}$ situations ($\geq$30\%).}
	\label{fig:coreshell-result}
\end{figure}

To further examine the influence of mechanical stresses on the reduction, the reduction degree, as well as the reduction degree rate, are shown in~\cref{fig:coreshell-result}. The reduction degree $f$ and its rate $\dot{f}$ are calculated as
\begin{equation}
\label{eq:reduction-degree}
f=1-\frac{\int_{\Omega}x_\mathrm{O}dV}{\int_{\Omega}x_\mathrm{O}^{\mathrm{init}}dV}\qquad\mathrm{and}\quad
\dot{f}=-\frac{\int_{\Omega}\dot{x}_\mathrm{O}dV}{\int_{\Omega}x_\mathrm{O}^{\mathrm{init}}dV},
\end{equation}
with $x_\mathrm{O}^{\mathrm{init}}$ being the initial oxygen content ($x_{w}$) within the sample.

The reduction degree and reduction degree rate curves in~\cref{fig:coreshell-reduction-degree,fig:coreshell-reduction-degree-rate} show a very similar shape at the beginning of the reaction for all cases.
This is attributed to the low-stress state at the beginning of the reaction.
However, as the reduction proceeds, different cases with different volume change ratio start to diverge due to the accumulated deformation and increased stress levels.
The non-monotonic behavior explained above becomes also visible.
The reduction degree curves for cases with a 10\% to 30\% transformation volume change are very close to the dashed line (zero volume change, i.e. no mechanical coupling), indicating a similar reduction rate.
However, the cases with 35\% and 40\% volume change reduction are significantly faster than the reference scenario without volume change (0\%).
As explained above, the occurrence of transformation induced high stresses (in the cases of the volume change fraction of $\alpha$-iron $\Omega_{f}=30\%$ and $\Omega_{f}=40\%$) results in a faster phase transformation. 
Thereby, a faster reaction kinetics is achieved.
Consequently, the sample with $\Omega_{f}=40\%$ shows a faster reduction reaction than the other cases.
The reduction degree rate under such a condition (blue dots in~\cref{fig:coreshell-reduction-degree-rate}) shows a higher plateau than the other cases.

\begin{figure}[H]
	\centering
	\includegraphics[width=0.75\linewidth]{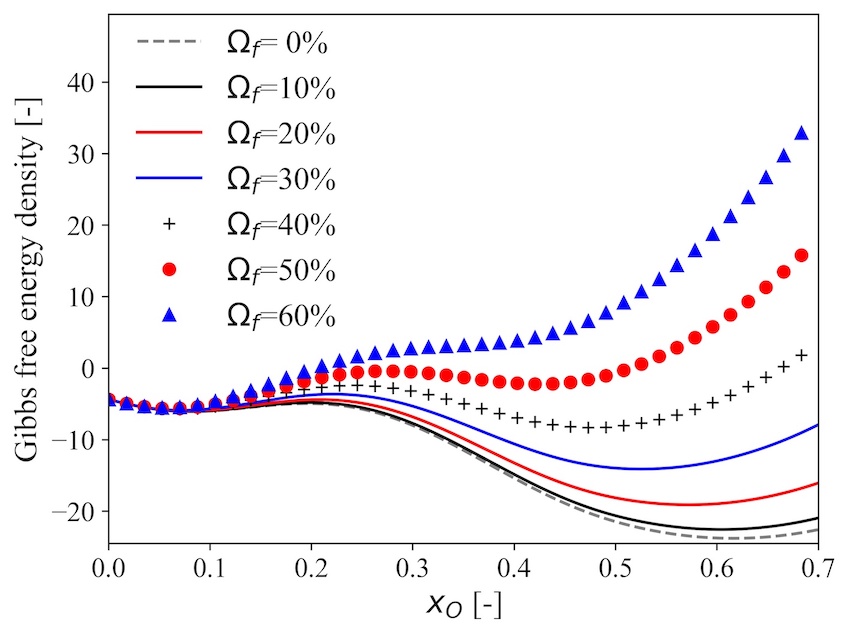}
	\caption{Mechanically coupled Gibbs free energy densities for different $\Omega_{f}$ values, where the solid/dashed lines with different colors indicate the coupling with small volume change ($\Omega_{f}$), while the markers represent the coupling with large volume change ($\Omega_{f}$).}
	\label{fig:mechanical-free-energy}
\end{figure}

To better understand the non-monotonic effect of the mechanical stresses, a simplified free energy density (shown in~\cref{fig:mechanical-free-energy}) with a 1D mechanical coupling is introduced as follows:
\begin{equation}
\label{eq:free-energy-mechanical}
f_\mathrm{fw}^{\mathrm{che}}(x_\mathrm{O}) =\frac{1}{2}E\Omega_{f}^{2}x_\mathrm{O}^{2}+\begin{cases}
f_{f}^{\mathrm{che}}(x_\mathrm{O}) & \text{if $x_\mathrm{O}\leq x_{f}$} \\
f_{f}^{\mathrm{che}}(x_\mathrm{O})(1-h(\frac{x_{O}-x_{f}}{x_{w}-x_{f}}))+f_{w}^{\mathrm{che}}(x_\mathrm{O})h(\frac{x_{O}-x_{f}}{x_{w}-x_{f}})&\text{if $x_{f}<x_\mathrm{O}\leq x_{w}$}\\
f_{w}^{\mathrm{che}}(x_\mathrm{O})&\text{otherwise}
\end{cases},
\end{equation}
where $E=220~\si{\giga\pascal}$ has been used for the plot as shown in~\cref{fig:mechanical-free-energy}. In this case, we assume that the material properties (the Young's modulus $E$ and the volume change $\Omega_{f}$ of $\alpha$-iron) are the same in all phases. The phase-dependent parameters are not taken into account.
It is worth noting that the couplings we have introduced here are based on~\cref{eq:free-energy-mechanical}, and not in the form of the fully coupled model described in~\cref{sec:model}.
~\cref{fig:mechanical-free-energy} shows that an increasing transformation volume change $\Omega_{f}$, from 0\% to 60\%, can result in a larger elastic energy contribution to the system, an effect that can retard the phase transformation.
However, once $\Omega_{f}$ is large enough, for instance, 40\% and 60\% in the current case, the contribution of the elastic energy becomes substantial: such a  large elastic energy contribution changes the shape of the free energy from a double-well system to a single-well system.
This means that a large mechanical contribution thermodynamically destabilizes the second phase completely and results in faster phase transformation.

\subsubsection{Reduction reaction in an elasto-plastic system}
As seen in the previous section, during the reduction reaction in an elastic system, the stress can for certain configurations reach such unrealistically large values as 22.4~\si{\giga\pascal}, as shown in~\cref{fig:coreshell-elastic-result} (right).
Such a huge accumulated elastic stress translates to a corresponding elastic energy contribution and can thus significantly modify the reduction kinetics, as discussed in~\cref{sec:stress-influence}.
Realistically, such stress levels are not reached in the material and inelastic relaxation phenomena such as plastic deformation, delamination, and damage evolution will limit the maximum elastic stress that is reached in the system.
Therefore, in this section, plastic deformation and its impact on the reduction reaction are investigated.
It should be noted that for simplicity, we only consider linear isotropic hardening here. The yield stress and hardening modulus for the reduction reaction at 700~\si{\degreeCelsius} are set to 300~\si{\mega\pascal} and 1.27~\si{\giga\pascal} for both solid phases (w\"ustite and $\alpha$-iron)~\cite{wang2013effect,wang2018mechanical}, respectively.
\begin{figure}[H]
\centering
\includegraphics[width=0.95\linewidth]{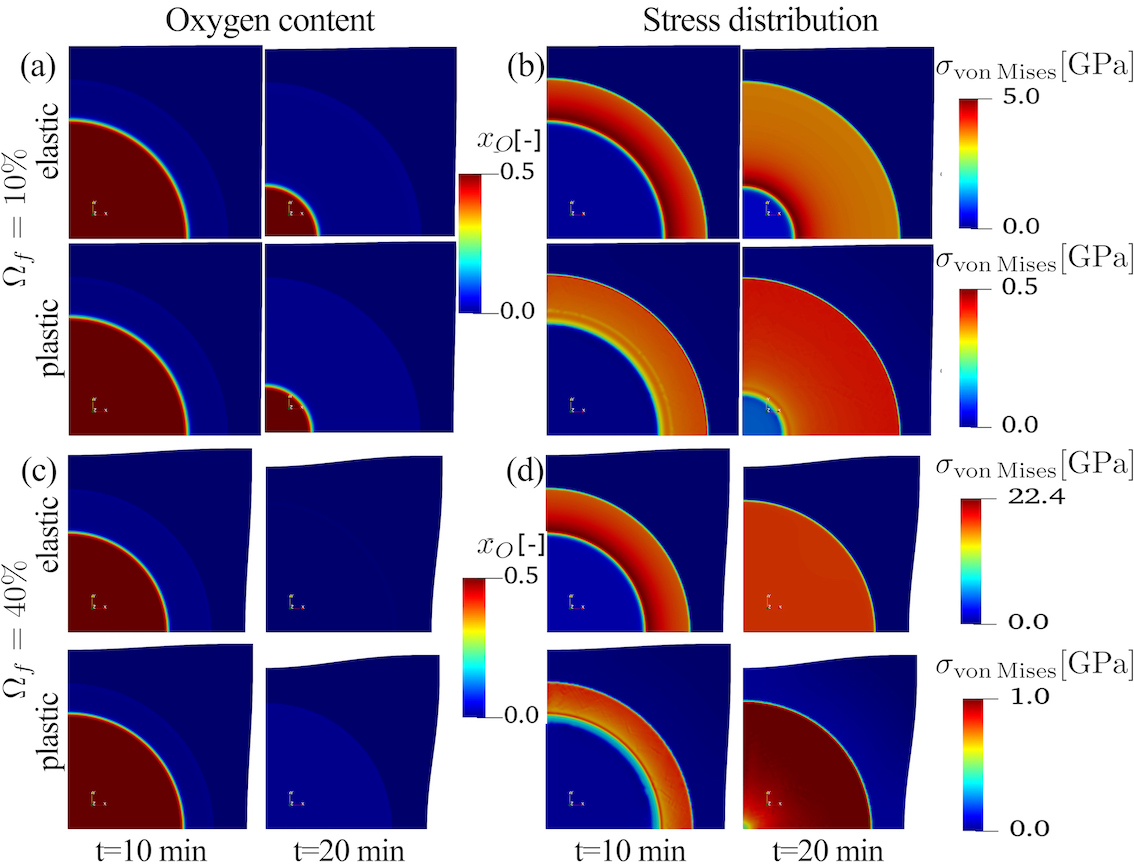}
\caption{Comparison between purely elastic deformation and elastoplastic deformation. (a) and (b) plot the oxygen molar fraction and von Mises stress for the samples with 10\% volume change at different times. (c) and (d) indicate the identical quantities for the samples with 40\% volume change at different times.}
\label{fig:coreshell-plastic-contour}
\end{figure}
~\cref{fig:coreshell-plastic-contour} plot the oxygen molar fraction and von Mises stress of the sample with different transformation volume changes at different reaction times.
As expected, considering the elasto-plastic material behaviour will result in lower retained elastic stress levels than in the case of purely elastic constitutive response as shown in~\cref{fig:coreshell-plastic-contour}(b) and (d).
In particular in the case of $\alpha$-iron phase's volume change of $\Omega_{f}=10\%$, the stress level remains relatively low for both, the elastic and the elasto-plastic constitutive case, especially during the early stages of the reaction (t=10~min). Therefore, the oxygen content distribution is almost identical in the two cases with (a) purely elastic and (b) elasto-plastic material response.
The stress level, however, will rise to 5.0~\si{\giga\pascal} as the reaction proceeds in the elastic case, while in the elastoplastic case the sample's stress only reaches 0.5~\si{\giga\pascal}, due to inelastic relaxation, i.e. onset of plasticity.

The differences between elastic and elasto-plastic material response become particularly apparent in the case of larger volume changes (i.e. $\Omega_{f}=40\%$), as shown in~\cref{fig:coreshell-plastic-contour}(c) and (d).
The von Mises stress in the purely elastic sample exceeds 22.4~\si{\giga\pascal} in large deformation situations, but the maximum value in the elasto-plastic sample is only 1.0~\si{\giga\pascal}.
As a result, the oxygen content distribution of these two samples already shows a difference at the early stage of the reaction (t=10~min).
As shown in the first row of~\cref{fig:coreshell-plastic-contour}(c), the purely elastic sample has very low oxygen content at t=20~min, whereas only a small amount of oxygen (light blue region) is present inside the center of the elasto-plastic sample.
Furthermore, the stress distribution in the purley elastic deformation case is nearly uniform as the core becomes small or even vanishes, as shown on the top-right of~\cref{fig:coreshell-plastic-contour}(d) when t=15~\si{\minute}.
Moreover, as $\Omega_{f}$ increases from 10\% to 40\%, the stress level in the elastoplastic cases does not increase as dramatically as the purely elastic case. 

\begin{figure}[H]
	\centering
	\begin{subfigure}{.49\textwidth}
		\centering
		\includegraphics[width=0.995\linewidth]{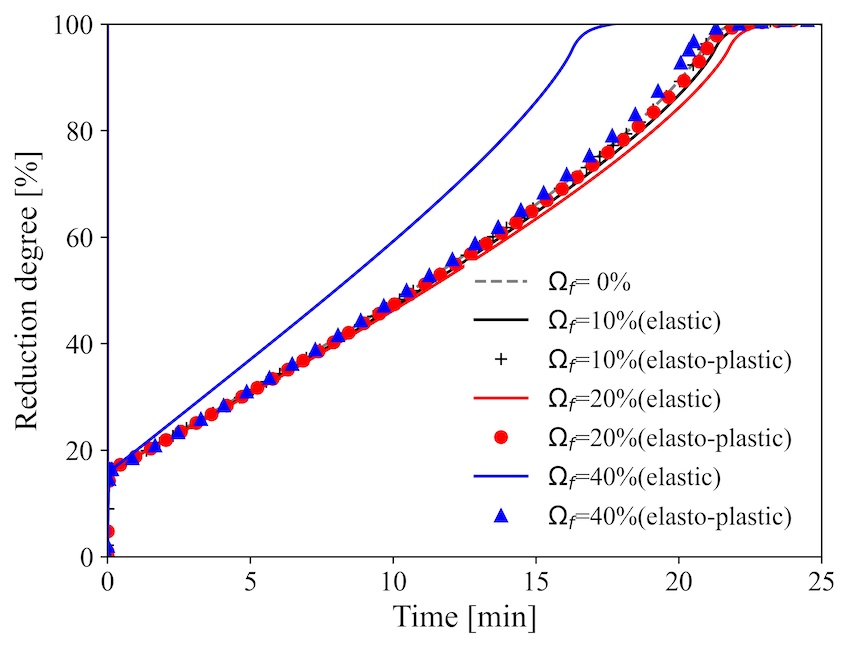}
		\caption{Reduction degree at 700~\si{\degreeCelsius}}
		\label{fig:coreshell-f}
	\end{subfigure}
	\begin{subfigure}{.49\textwidth}
		\centering
		\includegraphics[width=0.995\linewidth]{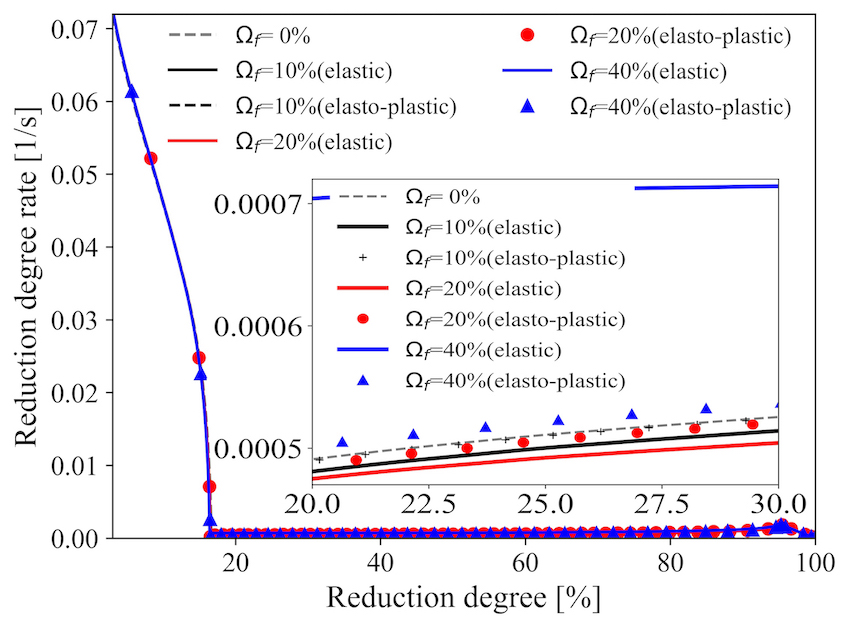}
		\caption{Reduction degree rate at 700~\si{\degreeCelsius}}
		\label{fig:coreshell-df}
	\end{subfigure}
	\caption{Reduction profile comparison between elastic and elasto-plastic deformation. (a) Reduction degree profiles for different cases and, (b) reduction degree rate profiles. The elasto-plastic deformation is represented by distinct colored markers, whereas the elastic deformation is represented by different colored solid lines.}
	\label{fig:coreshell-elastoplastic-result}
\end{figure}

The reduction degree and rate for the elastic and the elasto-plastic material response cases are plotted in~\cref{fig:coreshell-elastoplastic-result} to further investigate the plastic deformation effect on the reduction kinetics.

The reduction degree in the elasto-plastic case (different colored markers) follows the same patterns as the elastic case (solid lines with different colors) at the start of the reaction.
As the reaction proceeds, elasto-plastic samples with varying volume change ratios ($\Omega_{f}=10\%$, 20\%, and 40\%) show similar reduction degrees, as shown in~\cref{fig:coreshell-f}.
This is attributed to the elasto-plastic deformation, where the yield stress and hardening are much smaller than the bulk modulus in the purely elastic case.
This means that in the elasto-plastic material response case the local stress values are capped to moderate levels even under large volume change ratios, due to stress relaxation in the form of inelastic deformation.
Once the volume change ratio increases to the larger value of $\Omega_{f}=40\%$, the stress in the elasto-plastic case also reaches a higher level (about 1.0~\si{\giga\pascal}). However, this stress is still quite small compared to the purely elastic case with the same volume change ratio.
Also, once the yield stress for the onset of plastic deformation has been reached, the system stress relaxes and the elastic free energy contribution to the system is reduced.
The faster reaction kinetics is, therefore, more apparent for cases of higher volume change and for the purely elastic material response case. For elasto-plastic material response, regardless of the volume change, all cases behave similar to the case with no mechanical coupling ($\Omega_{f}=0\%$), due to the maintained stress level which is capped at the yield point.
In the later stages of the reduction ($f\geq$18\%), ~\cref{fig:coreshell-df} confirms that the reduction degree rate of the elasto-plastic sample is quite similar to that of the elastic sample with a lower $\Omega_{f}$ value.
The main conclusion of this section is thus, that compared with the purely elastic material response case, the plastic deformation limits the stress values in the system to the comparably moderate flows stress level, leaving only a limited influence of the remaining elastic stress on the overall reduction reaction.
One must of course consider here that this is true for the current simulation case, where the effects of the dislocations (which are the carriers of the inelastic material relaxation) on the mass transport and nucleation kinetics have not been explicitly considered.
This fact is similar to the purely elastic system with a low volume change fraction, where the mechanical stress hinders the transformation as explained in the previous section.
In contrast to the purely elastic case, in the elasto-plastic system the value of the volume change fraction ($\Omega_{f}$) has a minor impact on the system. 

\subsection{Simulation of an experimentally observed reduction scenario: the effects of real microstructures}
\begin{figure}[H]
	\centering
	\includegraphics[width=0.995\linewidth]{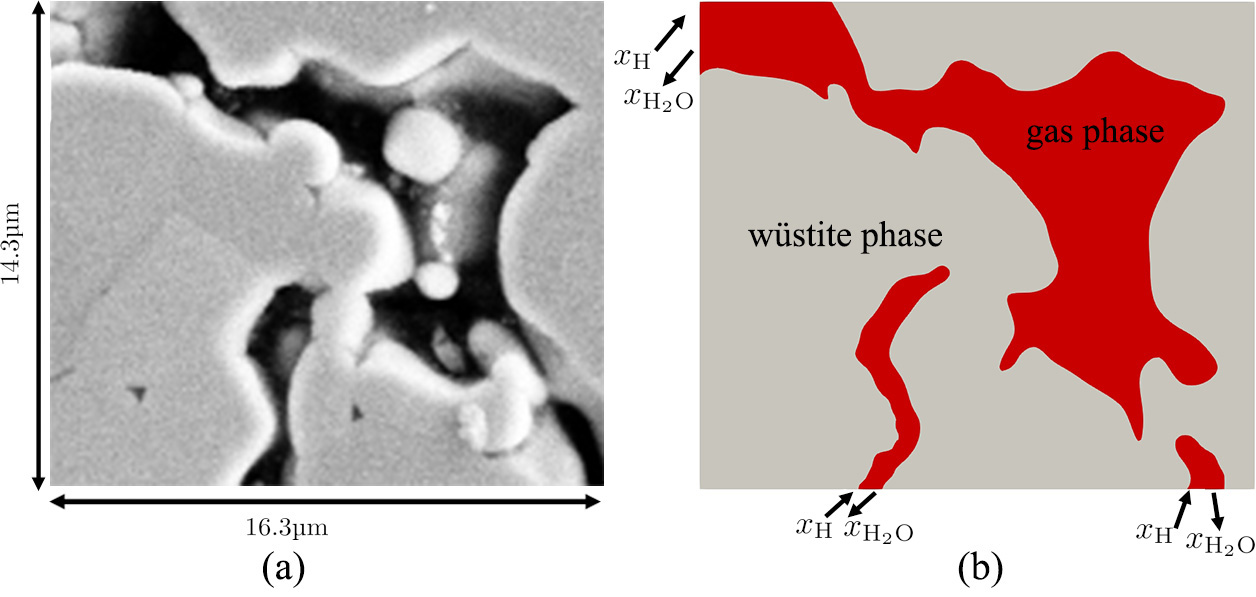}
	\caption{(a) SEM image of an originally hematitic pellet, which was subjected to hydrogen reduction, the last stage of which is the wüstite-to-iron transformation, (b) constructed model based on the SEM image. The gray color is the w\"ustite phase and the red part represents the gas phase. The arrows indicate the employed boundary conditions, namely, the ingress of hydrogen and the escape of the water.}
	\label{fig:microstructure-schematic}
\end{figure}

\begin{figure}[H]
	\centering
	\includegraphics[width=0.95\linewidth]{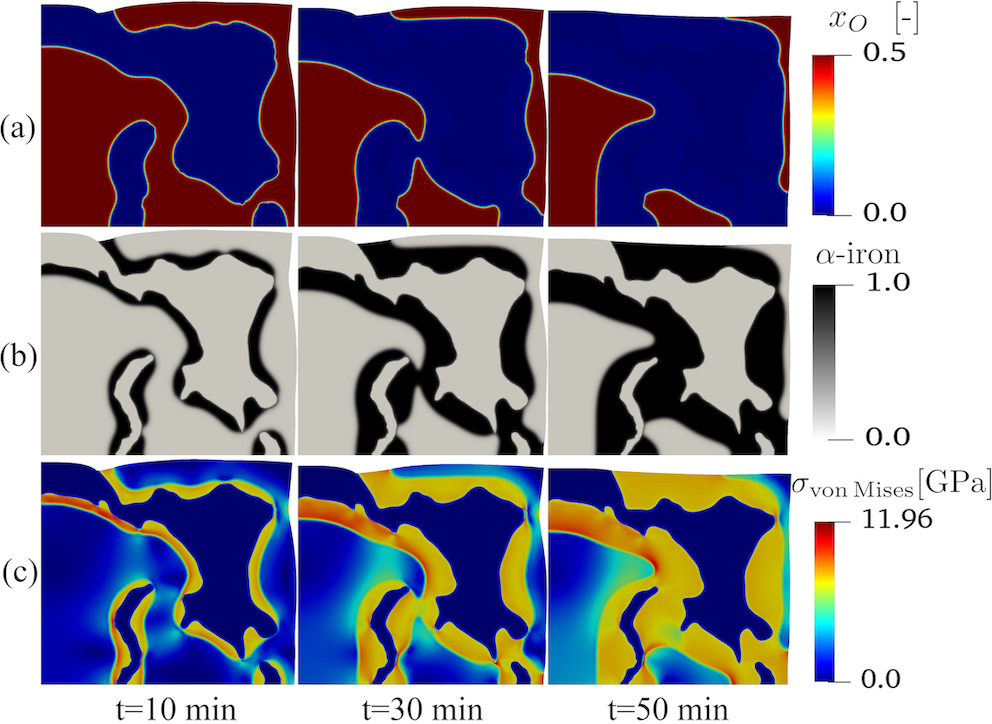}
	\caption{Contour plot of (a) oxygen molar fraction, (b) $\alpha$-iron phase order parameter, (c) von Mises stress for the purely elastic sample during reduction reaction.}
	\label{fig:microstructure-elastic-result}
\end{figure}

Real iron oxides typically have complex microstructures. On the one hand, they inherit complex defects structures from the mining, beneficiation, and pelletizing steps and on the other hand, they develop additional microstructure features during the reduction process, involving dislocations, interfaces, cracks, and pores~\cite{KIM2021116933}.
We therefore apply the fully coupled model in this section to the simulation of the reduction process using an experimentally determined microstructure as a starting material.
The data for the initial microstructure was obtained from an initially hematite pellet, which was subjected to a hydrogen reduction process during which the wüstite evolved~\cite{KIM2021116933}, i.e. we simulate here the last and most sluggish stage of the reduction process. 
For simplicity, we assume here that the w\"ustite islands fully inherit the morphology from the hematite islands in the direct reduction pellet.
The secondary pores and other microstructural defects (e.g. dislocations, interfaces) generated during the reduction of hematite to w\"ustite were neglected in this simulation approach. The microstructure of the initial sample was characterized using a Zeiss-Sigma 500 scanning electron microscope (SEM). 
\cref{fig:microstructure-schematic} depicts the geometry of the sample as well as the converted model with the applied boundary conditions.
The geometry information has been extracted from the SEM image using the Gaussian filters provided by the scikit-image package~\cite{scikit-image}.
At the edge of the gas phase channels, the chemical potentials of hydrogen and water have been fixed to simulate the ingress of more hydrogen and the escape of water during the reduction.
In the first approach, the fully coupled model is applied considering only elastic material behavior.
Furthermore, at the edges of the open channels in~\cref{fig:microstructure-schematic} (b), the hydrogen content is fixed to maintain a constant value of 1\%, whereas the molar fraction of water is set to be zero at the same edges to remove the generated water from the system. The volume change fraction of $\alpha$-iron ($\Omega_{f}$) is set to 20\% in this simulation.
As shown in~\cref{fig:microstructure-elastic-result} (c), the high stress values are located at the interfaces, where phase transformation takes place. The maximum von Mises stress could reach about 12~\si{\giga\pascal}.
The iron oxide region aligned along the solid-gas interface transformed first to the $\alpha$-iron phase, due to the high molar fraction of hydrogen intruding from the channel's edge, as shown in~\cref{fig:microstructure-elastic-result} (a) and (b).
Also, the right part of the sample has a higher volume fraction occupied by channels and pores than the left side, so that the former is faster reduced than the latter, as shown in the third column of~\cref{fig:microstructure-elastic-result}.

\begin{figure}[H]
	\centering
	\includegraphics[width=0.85\linewidth]{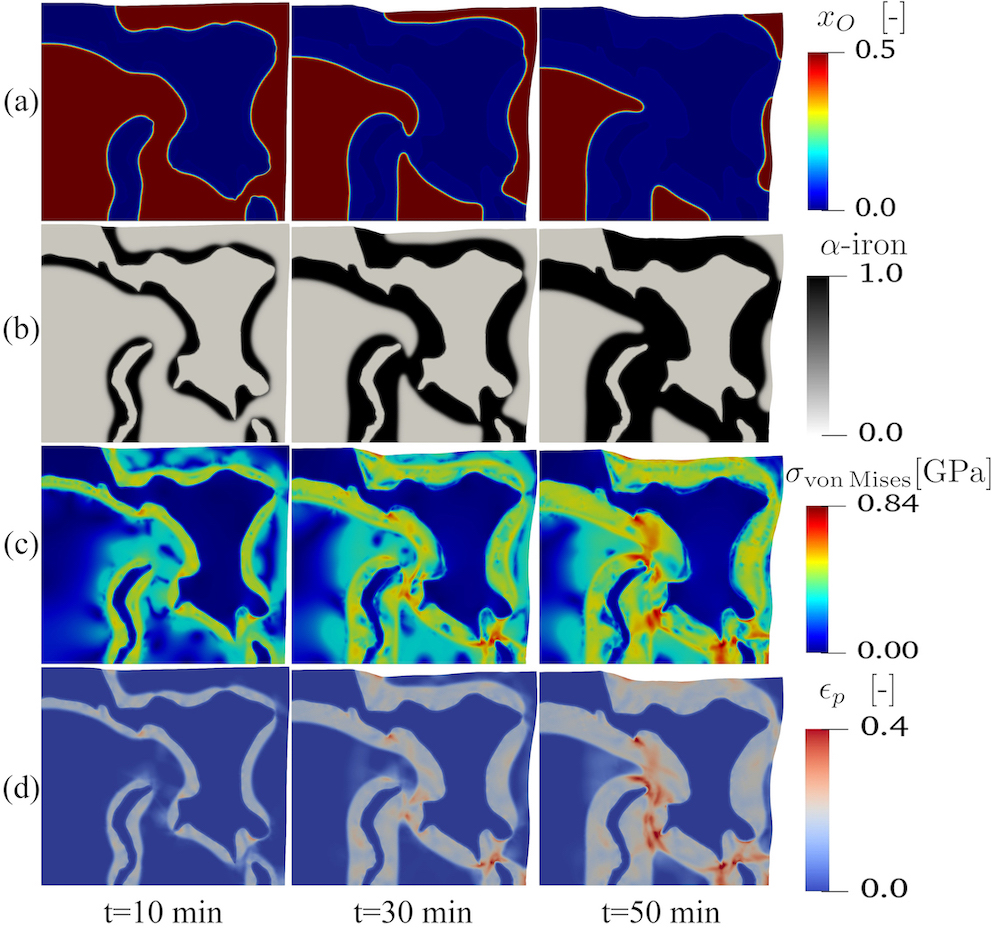}
	\caption{Contour plot of (a) oxygen molar fraction, (b) $\alpha$-iron phase order parameter, (c) von Mises stress, (d) effective plastic strain for the elasto-plastic constitutive solid response during the reduction reaction.}
	\label{fig:microstructure-plastic-result}
\end{figure}

In comparison, results for the (more realistic) case of the elasto-plastic material response for the same setup is shown in~\cref{fig:microstructure-plastic-result}.
In this case, the overall lower elastic stress level (due to plastic relaxation) results in a lower oxygen molar fraction and a higher degree of reduction than observed in the simulation conducted for the purely elastic case (~\cref{fig:microstructure-elastic-result})
However, rather than at the interface between the oxygen-poor phase ($\alpha$-iron) and oxygen-rich phase (w\"ustite phase), the highest stress value is built up within the $\alpha$-iron phase region, associated with the highest effective plastic strain, as shown in~\cref{fig:microstructure-plastic-result} (d).
As discussed in the previous section, the plastic strain is associated with a stress relaxation when the yield point is reached, i.e. the stress increases only very moderately with ongoing reduction.
This can cause a delaying effect of the stress on the phase transformation, which leads to a faster reduction than observed for the purely elastic simulation scenario.
As a result, the oxygen content within the sample is lower than in the purely elastic case at the same reaction time, as seen in ~\cref{fig:microstructure-elastic-result} (a) and ~\cref{fig:microstructure-plastic-result} (a).

As the reaction continues, for instance at t=30~min and t=50~min, the difference in the oxygen content and the $\alpha$-iron phase fraction between the purely elastic and elasto-plastic cases increases, as shown in ~\cref{fig:microstructure-plastic-result} (a) and (c).

\begin{figure}[H]
	\centering
	\includegraphics[width=0.85\linewidth]{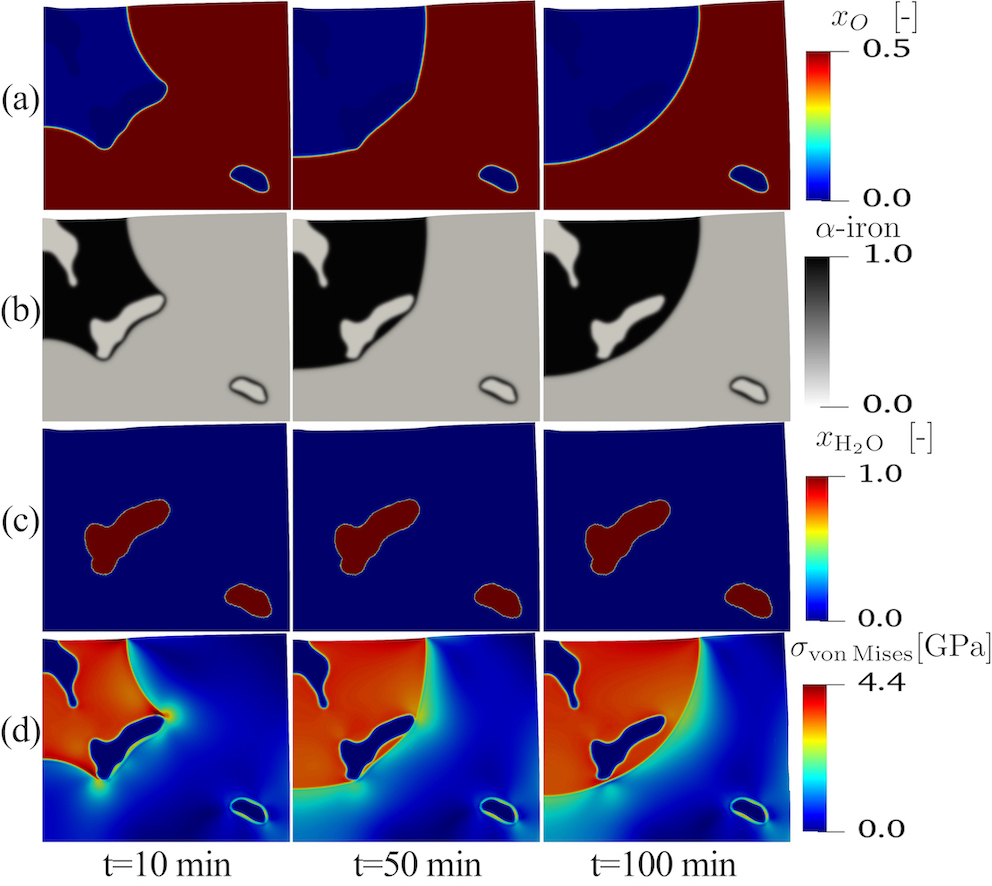}
	\caption{Analysis of a single pore scenario, for the case of elastic material response. Contour plot of (a) oxygen molar fraction, (b) $\alpha$-iron phase order parameter, (c) water molar fraction, (d) von Mises stress for the purely elastic constitutive solid response with two isolated pores during the reduction reaction.}
	\label{fig:microstructure1-elastic-result}
\end{figure}

To investigate water formation during the reduction reaction, a sample with two isolated pores has been initialized, as shown in~\cref{fig:microstructure1-elastic-result}.
It is worth noting that the reduction reaction will come to a halt once the isolated pores are completely filled with water, namely $x_{\ce{H2O}}=1$.
Consequently, the reaction supply rate $r$ in~\cref{equ:EmpModRea} is treated as zero in such a case.
\begin{figure}[H]
	\centering
	\includegraphics[width=0.85\linewidth]{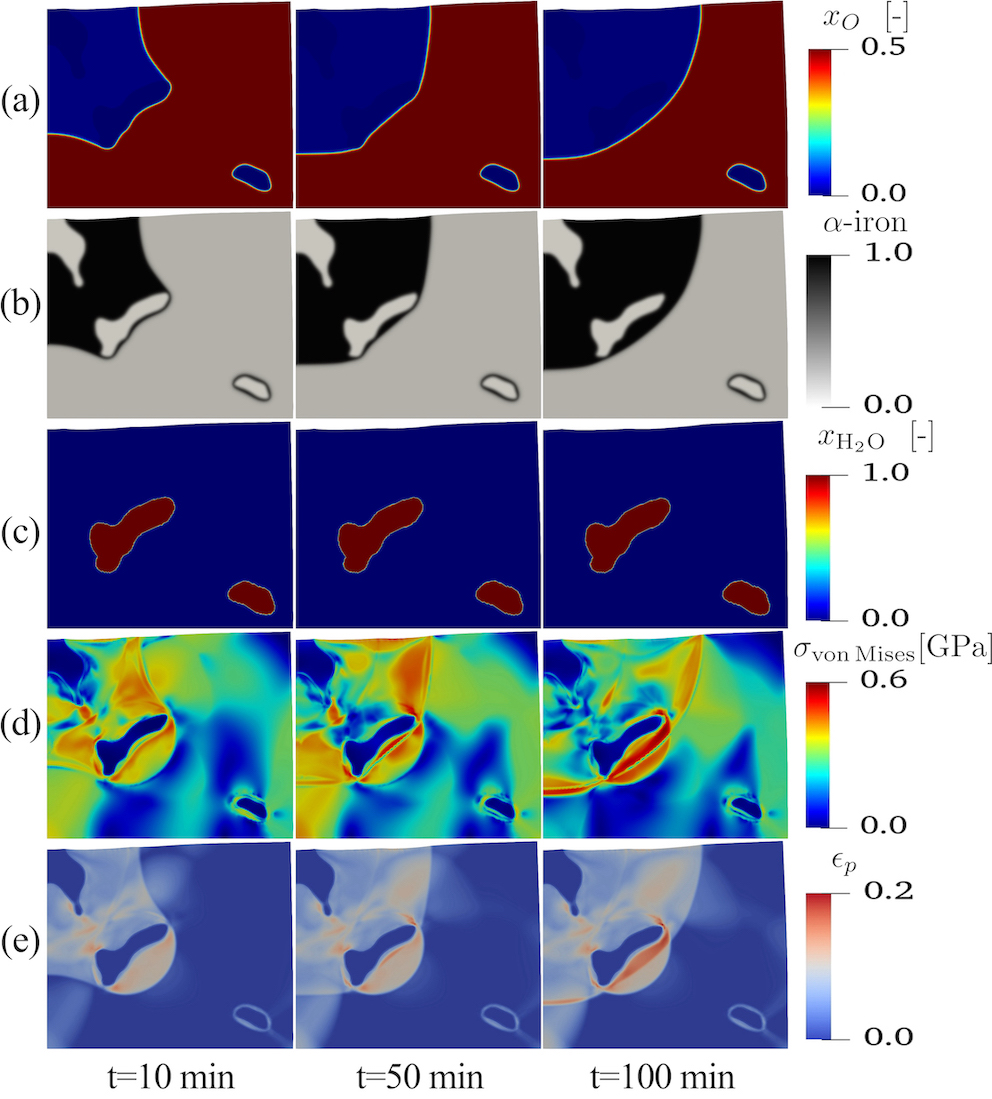}
	\caption{Analysis of the same single pore scenario as in~\cref{fig:microstructure1-elastic-result}, however, for the elasto-plastic material response. Contour plot of (a) oxygen molar fraction, (b) $\alpha$-iron phase order parameter, (c) water molar fraction, (d) von Mises stress, (e) effective plastic strain for the case of elasto-plastic constitutive response with two isolated pores during the reduction reaction.}
	\label{fig:microstructure1-plastic-result}
\end{figure}
As illustrated in~\cref{fig:microstructure1-elastic-result} (b) and~\cref{fig:microstructure1-plastic-result} (b), the $\alpha$-iron phase is produced first along the solid-gas interface.
Due to the low stress at the early stages of the reduction, the simulation with the elasto-plastic material response in~\cref{fig:microstructure1-plastic-result} (a) has a similar oxygen molar fraction as the case with purely elastic response in~\cref{fig:microstructure1-elastic-result} (a).
Since the hydrogen must diffuse through the w\"ustite phase to reach the isolated pore, only a limited quantity of $\alpha$-iron phase is created near the surface of the isolated pores, which is far away from the open gas channel.
This is also confirmed from~\cref{fig:microstructure1-elastic-result} (c) and~\cref{fig:microstructure1-plastic-result} (c), where the isolated pores are completely filled up with water.
Furthermore, as seen in~\cref{fig:microstructure1-elastic-result} (c) and~\cref{fig:microstructure1-plastic-result} (c), the molar fraction of water in channels connecting to the sample's edge is nearly zero in all the snapshots, indicating that the water has been entirely removed from the reaction zone.
Since water's diffusivity in the solid phase is nearly negligible, once the isolated pores are filled with water, the reduction reaction stops locally around the pore as seen from~\cref{fig:microstructure1-elastic-result} (c) and~\cref{fig:microstructure1-plastic-result} (c). During the reduction process, the $\alpha$-iron phase is relatively low near these pores.
Unlike for the case of the elastic material response shown in~\cref{fig:microstructure1-elastic-result} (d), the elasto-plastic reaction can result in larger stresses around the surface of isolated pores than the point within the solid phase which is far away from the pores' surface.
However, the maximum stress level is still observed in the $\alpha$-iron phase around the interconnected channels, as shown in~\cref{fig:microstructure1-plastic-result} (d).
This result demonstrates that the reaction around the isolated pores occurs slower than the reaction close to connected channels.
Despite the fact that the stress levels in these two samples are considerably different, the trapped water can slow down the reaction significantly.
As a result, the reaction is limited to the interface surrounding the open channels.
Investigating the effects of mechanical deformation on damage development and porosity evolution represent the work in progress to be reported in follow-up studies. 

\section{Conclusions}
\label{sec:conclude}
We introduced, tested, and applied a chemo-mechanically coupled phase-field (PF) model to study the iron oxide direct reduction with  gaseous hydrogen.
The constitutive laws for the diffusion of oxygen, the phase transformation from w\"ustite to $\alpha$-iron, as well as the elasto-plastic deformation have been derived from the system free energy.
The model makes use of an existing thermodynamic database for the oxygen-dependent free energies of w\"ustite and $\alpha$-iron.
In particular, the thermodynamic database for the oxygen-dependent free energies of w\"ustite and $\alpha$-iron has been incorporated within this PF model.

We have first benchmarked our model for an oxygen-dependent free energy scenario in a rectangular domain. Simulation results show that the predicted phase fractions in the equilibrium state agree very well with the analytical results.
Next, an iron oxide sample with a core-shell structure, where the shell consists of a freshly reduced iron layer and the core of wüstite oxide, has been examined.
We find that the high volume change from w\"ustite to $\alpha$-iron can result in very high stress (of the order of tens of~\si{\giga\pascal}) for the (more academic) case of purely elastic material response.
We identified two regimes for a scenario with the purely elastic material response, mainly governed by the volume change between w\"ustite and $\alpha$-iron.
At relatively moderate volume changes below 25\%, the accumulated elastic stress that builds up during the reduction will slightly slow down the reduction kinetics.
However, for higher transformation volume changes (above 25\%), the accumulated stress during the transformation and the resulting elastic energy have a substantial effect on the shape of the total free energy of the system.
This additional stored elastic energy (for the reference case of a material with purely elastic response) thus creates a substantial additional driving force which accelerates the phase transformation and results in overall faster reaction kinetics.

This effect relaxes when considering also plastic deformation: the  high maximum von Mises stress of  22.4~\si{\giga\pascal} observed for the purely elastic case drops to the much lower yield stress level of only  1.0~\si{\giga\pascal} for the same volume change ratio for the elasto-plastic case.
As a consequence, the reduction degree and rate predicted for the elasto-plastic material shows similar patterns as for the purely elastic cases at small volume change ratios.
It should be noted that in the current work the effect of plasticity on the reduction reaction is only coupled through the associated mechanical energy density, which is much larger for the unrelaxed purely elastic case than for the elasto-plastic case, where the energy density is capped at the respective yield points.
A more realistic coupling should therefore also include kinetically relevant effects that come with the presence of dislocations and cracks etc., such as the multiple effects associated with the presence of such lattice defects on the transport and surface reaction dynamics.
These higher-order effects will be investigated in future work.
We conducted further simulations on experimentally observed microstructures and observe a significant role of open channels and pores during the reduction reaction.
It is shown that the formation of the $\alpha$-iron phase highly depends on the availability of local free surface areas, provided through the local channel and porosity features, acting both, though the change in the local stress state as well as accelerated material transport.

In summary, we demonstrate that the stresses that build up inside of the iron oxide during hydrogen-based direct reduction for both, elastic and elasto-plastic deformation scenarios can have a significant effect on the transformation behaviour, oxygen diffusion, reduction kinetics, and metallization.
Furthermore, we show that the microstructure plays an important role in the reaction kinetics.
Including information about the connectivity of the pores and channels is crucial for an accurate prediction of the reduction dynamics.
In future work, therefore, the evolution of the local porosity and delamination features due to loss of oxygen and due to mechanical stress during the reduction reaction will be incorporated into the model. 

\section*{Acknowledgements}
The authors gratefully acknowledge the computing time granted by the Paderborn Center for Parallel Computing (PC$^{2}$).
Dr. Yan Ma acknowledges financial support through Walter Benjamin Programme of the Deutsche Forschungsgemeinschaft (Project No. 468209039). 

\bibliography{mybibfile}

\end{document}